\documentclass[12pt,preprint]{aastex}

\newcommand{\gcc}{\mathrm{g~cm^{-3} }}
\newcommand{\cms}{\mathrm{cm~s^{-1}}}

\slugcomment{in preparation}

\usepackage{epsf,color,dcolumn}

\newcolumntype{d}{D{.}{.}{-1}}

\epsfverbosetrue

\begin{document} 

\title{Direct Numerical Simulations of Type Ia Supernovae Flames II: The
Rayleigh-Taylor Instability}

\shorttitle{R-T Unstable Flames in Type Ia SNe}
\shortauthors{Bell et al.}

\author{J.~B. Bell\altaffilmark{1},
        M.~S. Day\altaffilmark{1},
        C.~A. Rendleman\altaffilmark{1},
        S.~E. Woosley\altaffilmark{2},
	M.    Zingale\altaffilmark{2}}

\altaffiltext{1}{Center for Computational Science and Engineering,
                 Lawrence Berkeley National Laboratory,
                 Berkeley, CA 94720}

\altaffiltext{2}{Dept. of Astronomy \& Astrophysics,
                 The University of California, Santa Cruz,
                 Santa Cruz, CA 95064}

\begin{abstract}

A Type~Ia supernova explosion likely begins as a nuclear runaway near
the center of a carbon-oxygen white dwarf.  The outward propagating
flame is unstable to the Landau-Darrieus, Rayleigh-Taylor, and
Kelvin-Helmholtz instabilities, which serve to accelerate it to a
large fraction of the speed of sound.  We investigate the
Rayleigh-Taylor unstable flame at the transition from the flamelet
regime to the distributed-burning regime, around densities of
$10^7~\gcc$, through detailed, fully resolved simulations.  A low Mach
number, adaptive mesh hydrodynamics code is used to achieve the
necessary resolution and long time scales. As the density is varied,
we see a fundamental change in the character of the burning---at the
low end of the density range the Rayleigh-Taylor instability dominates
the burning, whereas at the high end the burning suppresses the
instability.  In all cases, significant acceleration of the flame is
observed, limited only by the size of the domain we are able to study.
We discuss the implications of these results on the potential for a
deflagration to detonation transition.
\end{abstract}


\section{INTRODUCTION}
\label{sec:intro}

Accelerating a thermonuclear flame to a large fraction of the speed of
sound (possibly supersonic) is one of the main difficulties in
modeling Type~Ia supernovae (see the recent review by
\citealt{hillebrandtniemeyer2000} and references therein).  Numerical
results have shown that a prompt detonation does not yield the right
abundance pattern to account for the observations \citep{arnett1971};
consequently, the burning must begin as a subsonic flame.  One
dimensional simulations show that the flame needs to reach
approximately one third
of the speed of sound to properly account for the explosion energetics
and nucleosynthetic yields \citep{woosley84,nomoto84}, although, a
transition from a deflagration to a detonation at some late stage of
the explosion \citep{khokhlov1991,niemeyerwoosley1997} is not ruled
out.  Laminar flame speeds are too slow by several orders of magnitude
\citep{timmeswoosley1992}, so instabilities and the interaction with
flame-generated turbulence are turned to in hopes of providing a
mechanism to considerably accelerate the flame.

In this paper, we focus on the effects of the Rayleigh-Taylor (RT)
instability \citep{taylor1950, chandra} on the flame.  After ignition,
the subsonic flame moves outward from the center of the star.  The
pressure is essentially continuous across the flame. Furthermore,
since the star has time to expand in response to the burning, this
pressure also remains constant in time.  As the flame propagates, it
leaves behind hot ash that is less dense than the cool fuel.  Gravity
points toward the center of the star, so this flame front is RT
unstable.  Bubbles of hot ash will rise and try to exchange places
with spikes of cool fuel, increasing the surface area of the flame and
leading to an enhancement of the flame speed.  It is possible that this
acceleration, operating on the stellar scale, can bring the flame
speed up to a significant fraction of the speed of sound.
Furthermore, the shear layer between the fuel and ash which develops
as the instability evolves is Kelvin-Helmholtz unstable, creating
turbulence in the region of the flame.  As the flame surface is
wrinkled by the RT instability, it will interact with this turbulence.
We look in detail at the interaction between the RT instability and
the flame on small scales through 2-dimensional spatially resolved
simulations in conditions appropriate to the late stages of a Type~Ia
supernova explosion.

The difference in scale between the white dwarf and the flame width is
enormous (up to 12 orders of magnitude), making direct numerical
simulations of the whole explosion process impossible.  Nevertheless,
simulations of RT unstable flames have been attempted for some time,
both on the full stellar scale and as smaller, constant density
micro-physical studies.  Various techniques and approximations are
used to follow an unresolved flame on these scales.  Some of the
earliest such calculations \citep{mullerarnett1982,mullerarnett1986}
used donor-cell advection to follow a combustion front through the
star, burning most of it, and releasing enough energy to produce an
explosion.  In their model, the flame was propagated essentially
through numerical diffusion.  The RT instability was resolved only on
the very largest scales of the star, missing the turbulence generated
on the small scales.  \citet{Livne1993} used a two-dimensional
implicit Lagrangian-remap hydrodynamics code with a mixing-length
subgrid model to follow a RT unstable flame.  The diffusion of
reactants in the Lagrangian-remap scheme is much lower than that used
in \citet{mullerarnett1982,mullerarnett1986}, so a more physical flame
speed could be used.  The flame speed was chosen to be either the
conductive (laminar) or turbulent speed---which ever was larger.
Again, the whole star was modeled, with coarse resolution, showing
that the RT instability sets up and greatly convolutes the flame
front, but the resulting acceleration left the flame too slow to match
observations.  \citet{Livne1993} suggested that more resolution is
needed to get the small scale physics right.

The effects of turbulence on the burning were investigated by
\cite{niemeyerhillebrandt1995b,khokhlov1995}.  Two-dimensional, small
scale RT simulations using a thickened-flame model, coupled to an
explicit Lagrange-remap PPM \citep{ppm} implementation to model a thin
flame front were performed by \citet{khokhlov1993} and later extended
to 3-d \citep{khokhlov1994}.  In this model, a reaction-diffusion
equation for the fuel concentration is solved along with the Euler
equations, with the conductivity and reaction rate modified to yield
the desired laminar flame speed and a flame that was a few
computational zones wide.  The thermal structure of the true flame is
stretched greatly, and, as a result, curvature effects of the flame
are not properly included.  Energy is deposited behind the flame to
couple this reaction-diffusion model to the hydrodynamics.  Similar
techniques have also been used for terrestrial flames (see for example
\citealt{orourke1979}).  Simulations of the RT instability at a
density of $10^8~\gcc$ lead the author to conclude that while the
flame accelerates, it is too small of an acceleration (when scaled up
to the size of the star) to account for the observations, and that a
deflagration-detonation transition may be more promising.  Owing to
the difference in behavior between two-dimensional and
three-dimensional turbulence, \citet{khokhlov1994} concludes that the
turbulence was more effective in increasing the flame speed in 3-d.
\citet{khokhlov1995} demonstrated scaling relations for the 3-d RT
unstable flame and concluded that the turbulent flame speed does not
depend on the small scale burning or the laminar flame speed.  

The independence of the turbulent flame speed on the laminar speed was
also postulated by \citet{niemeyerhillebrandt1995b}, who performed a
full-stellar model based on Kolmogorov scaling of turbulence from the
large stellar scales down to the flame scales.  Here, a turbulent
scaling model using the Gibson length (the scale at which an eddy is
completely burned during a single turnover) as the small scale cutoff
yields a turbulent flame velocity prediction for the grid scale.  This
was coupled with a subgrid model for the turbulent kinetic energy, and
implemented within the PPM algorithm.  Their simulations of the flame
propagating through the whole star showed the RT instability
dominating the flow, but the flame did not accelerate enough to
produce an explosion.

More recently, large scale simulations of pure deflagration explosion
have been performed using two different methods, level-sets
\citep{reinecke1999b,reinecke1999a,reinecke2002b} and a
thickened-flame model \citep{gamezo2003} with a Sharp-Wheeler
\citep{sharp1984,glimmli1988} RT subgrid model.  Both groups using PPM
as their underlying explicit hydrodynamics method.  Different numbers
and locations of the ignition points are used, but the 3-d simulations
by both groups release enough energy to explode the star, with the
burning proceeding solely as a deflagration.

These studies have demonstrated the tremendous effect the RT
instability has as the dominant acceleration mechanism of the
flame---to the point where, in some simulations, a deflagration alone
can unbind the star.  However, in all the cases discussed above, some
model for the flame or turbulence on scales smaller than the grid
resolution was required.  In this paper, we present fully resolved
simulations of RT unstable flames at the densities appropriate to the
late stages of the explosion---no flame model is used.  Fully
resolving the flame means that the effects of curvature and strain on
the flame are implicitly accounted for.  Curvature effects on flames
in conditions close to those studied here were found to be important
in \citet{flame-curvature}, where laminar flames were propagated in
diverging and converging spherical geometries.  Here, the effects can
be even larger as the flame will experience sharp kinks as it is
distorted by the RT instability.  To achieve converged steady-state
laminar flames, approximately five grid point resolution in the thermal
width is required.  This limits the size of the domain that can be
modeled significantly (although the adaptive mesh refinement employed
here helps greatly), but frees us from the need to deal with subgrid
models.  Resolved simulations at the smallest scales complement the
full star flame-model calculations, and it is hoped that progress from
the small scales up will allow us in the future to develop and
calibrate more accurate subgrid models.

Fully resolved multidimensional Type~Ia-like flame simulations are
rare \citep{niemeyerhillebrandt1995a, flamevortex, SNld}, mainly because the
flame moves so slowly, requiring excessive numbers of timesteps for
compressible codes.  Resolved calculations of the reactive RT
instability have been performed before \citep{vladimirova2003}, but
using model flames (a KPP-type reaction, \citealt{kpp}) and a
Boussinesq approximation, under conditions that are not directly
applicable to Type~Ia SNe.  When extrapolated to the astrophysical
regime, they predict a flame speed independent of the laminar speed,
similar to that found by \citet{khokhlov1995}.  However,
compressibility was neglected in these calculations.  Here we use a
low Mach number formulation of the equations of motion, which retains
the compressibility, but frees us from the restrictive acoustic timestep
constraint, and we use realistic input physics (reaction rate, EOS,
etc.)  appropriate to the conditions in a Type~Ia SNe explosion.

Flame instabilities stretch and wrinkle the flame, increasing the bulk
burning rate.  The SNe flame is subject to the Landau-Darrieus
\citep{landau:1944, darrieus:1938} and RT instabilities, as well as
the interaction with turbulence.  In a previous paper,
(\citealt{SNld}, henceforth referred to as paper~I), we looked at
Landau-Darrieus unstable flames, and validated the low Mach number
method for astrophysical flames, confirming the results of earlier
studies that the acceleration of the laminar flame is quite small,
while finding no evidence for the breakdown of the cusps in the
non-linear regime.  In the present paper, we look in detail at the
interaction between the burning and the RT instability.  A RT driven
flame behaves very differently from its laminar counterpart.  Aside
from looking at the increase in velocity over the laminar speed the
wrinkling provides, we seek to determine whether there is a simple
scaling relation that matches these results, and whether such a
relation holds over the whole range of densities, where the character
of the burning is expected to change dramatically.

In \S\ref{sec:distributedburning}, we discuss the effects of the
burning on the RT instability when we enter the distributed burning
regime.  In \S\ref{sec:numerics}, the low Mach number numerical method
and input physics are described, followed by the results of our
simulations in \S\ref{sec:results}.  We conclude in
\S\ref{sec:detonation} by discussing the implications of the RT
burning process on the explosion mechanism for Type~Ia SNe.

\section{\label{sec:distributedburning} THE EFFECTS OF BURNING ON THE RAYLEIGH-TAYLOR INSTABILITY}
\label{sec:rt}

The RT instability has been well studied analytically
\citep{taylor1950, layzer1955, chandra} and numerically, both as an
incompressible (see recent results by \citealt{young2001,cook2001})
and compressible (see for example \citealt{glimmli1988,
flash-validation}) fluid.  In the absence of burning, the growth of
the multimode RT instability in the non-linear regime is
characterized by two parameters, one describing the terminal velocity,
and the other describing bubble merger \citep{glimmli1988}.  Bubbles
are assumed to merge when the difference in their heights exceeds some
critical multiple of the smaller bubbles size.  This leads to an
expression for the average position of the interface,
\begin{equation}
\label{eq:sw}
h = \alpha A g t^2 \enskip ,
\end{equation}
\citep{sharp1984} where $\alpha$ is a function of the two parameters,
and $A =
(\rho_\mathrm{fuel}-\rho_\mathrm{ash})/(\rho_\mathrm{fuel}+\rho_\mathrm{ash})$~is
the Atwood number.  Much theoretical, computational, and experimental
work aims to measure $\alpha$, and some evidence suggests that it may
be constant for a wide range of initial conditions \citep{george2002}.
The velocity of the merged bubbles can be written as
\begin{equation}
\label{eq:sw2}
v =  c \sqrt{A g h} \enskip ,
\end{equation}
where $c \sim 1/2$ \citep{daviestaylor1950}.  These expressions were
all formulated for the non-reactive RT instability and studies of
rising bubbles.  Numerical experiments of reactive RT instabilities
in SNe conditions \citep{khokhlov1995} using a thickened-flame model
found agreement with the above velocity expression, with $c = 0.5$.
Expressions of this form have been used as a subgrid model for the
flame speed in full-star, pure deflagration calculations of Type~Ia
explosions in 1-d \citep{niemeyerwoosley1997} and 3-d
\citep{gamezo2003}, resulting in a successful explosion of the star.
These results suggest that the RT instability is the dominant
acceleration mechanism for the flame.

\citet{woosley1990} and \citet{timmeswoosley1992} argued that, unlike
a non-reactive RT instability, where all wavelengths can grow (in the
absence of surface tension), burning introduces a small scale cutoff
to the instability.  The growth rate for the pure RT instability is
determined by the dispersion relation
\begin{equation}
\omega^2 = g k A
\end{equation}
\citep{chandra}, where $\omega$ is the angular frequency and $k$ is
the horizontal wave number for the dominant mode.  We note that we
have ignored contributions due to surface tension.  By equating the
period for the growth of the instability to the laminar flame
propagation timescale, \citet{timmeswoosley1992} showed that there is
a critical wavelength below which any perturbations will be washed out
by burning.  This wavelength is called the fire-polishing wavelength,
$\lambda_\mathrm{fp}$:
\begin{equation}
\label{eq:firepolishing}
\lambda_\mathrm{fp} = 4 \pi \frac{v_{\mathrm{laminar}}^2}{g_{\mathrm{eff}}} \enskip ,
\end{equation}
where $g_{\mathrm{eff}} = g A $ is the effective gravitational
acceleration.  We note that this is a simple estimate, nevertheless,
it provides a useful measure of the effectiveness of the RT
instability on disrupting the flame.
Table~\ref{table:flameproperties} lists the flame properties, as
computed by the code described in \S \ref{sec:numerics}, including
$\lambda_\mathrm{fp}$ for the densities we are considering.

Stratification effects are negligible on the scales that we are
considering, since we are modeling a tiny fraction of the star.  The
relative change in pressure over our domain is
\begin{equation}
\frac{\Delta p}{p} = \frac{\rho g \Delta r}{p} \sim \frac{v_{\mathrm{SW}}^2}{c_s^2} = M_{\mathrm{SW}}^2 \ll 1 \enskip ,
\end{equation}
where we have expressed this ratio in terms of the Mach number of
the Sharp-Wheeler velocity for a RT flame encompassing our entire domain.
Since we are resolving the structure of the flame, the domain size
restrictions limit the speed-ups we see to about 10~times the laminar
speed, which gives a maximum $\Delta p/p$ of $O(10^{-8})$.  
 
As the density of the white dwarf decreases, the flame speed decreases
sharply while the flame width, $l_f$, increases.  As
Table~\ref{table:flameproperties} shows, at the densities we are
considering, we pass from the regime where $\lambda_\mathrm{fp} \ll l_f$ (low
densities) to $\lambda_\mathrm{fp} \gg l_f$ (high densities).  When
$\lambda_\mathrm{fp} \ll l_f$, the RT instability will dominate the
flame, to the point where it may no longer be continuous, and one can
no longer draw a simple curve from one side of the domain to the other
marking the flame position.  This increase in complexity of the flame
surface signifies that we are leaving the flamelet regime.

\citet{niemeyerwoosley1997} made estimates of the density where the
burning transitions from the flamelet regime to the distributed
burning regime, in which a distinguished local flame surface is not
apparent.  Using the flame properties tabulated in
\citet{timmeswoosley1992} and equating the Gibson scale to the flame
width, they found that for a 0.5~$^{12}$C\slash 0.5~$^{16}$O flame,
the burning will enter the distributed burning regime at a density of
$1$--$5\times 10^7~\gcc$.  \citet{niemeyerkerstein1997} refined this
estimate by suggesting that since the turbulence present in a Type~Ia
explosion is generated by the RT instability, it should follow
Bolgiano-Obukhov rather than Kolmogorov statistics.  They found a
slightly lower transition density of $O(10^7)~\gcc$.  At these
densities, the turbulence is intense enough to disrupt the flame.  We
note that this transition density is right in the range where the fire
polishing length becomes equal to the flame width, as discussed above.

This density range is also interesting in the context of deflagration
to detonation transitions (DDT).  Arguments based on nucleosynthesis
and one-dimensional modeling suggest that if the burning in a Type~Ia
supernova were to transition from a deflagration to a detonation, this
transition would need to occur at a density around $10^7~\gcc$ (see
for instance \citealt{hoeflichkhokhlov1996}).  The correlation between
this DDT density and the density marking the transition to the
distributed burning regime was pointed out by
\citet{niemeyerwoosley1997} and \citet{khokhlov1997}.  One-dimensional
turbulence studies of the flame-turbulence interaction at these
densities were performed by \citet{lisewski2000}, where it was
observed that the turbulence disrupts the flame sufficiently in
some cases to create local explosions as the hot ash comes in contact
with pockets of cool fuel.  If these regions of fuel are large enough,
these local explosions may be able to initiate a DDT.

Direct numerical simulations of thermonuclear flames at these
densities do not exist.  \citet{nbr1999} performed simplified flame model
calculations in 3-d using a source term relevant to flames in Type~Ia
supernovae.  They concluded that for densities larger than
$O(10^7)~\gcc$, scaling the flame speed in direct proportion to the
increase in surface area is a valid subgrid model.  For densities
below $10^7~\gcc$ however, they state that the effects of turbulence
become important.

As Table~\ref{table:flameproperties} shows, the laminar flame speed in
our density range is extremely subsonic---our highest laminar flame
Mach number is $2\times 10^{-5}$.  Simulating a flame at these
densities with a fully compressible code would be prohibitively
expensive, requiring millions of timesteps.  This in turn leads to a
large accumulation of error.  Instead, a low Mach number hydrodynamics
code \citep{SNeCodePaper} is employed for the present study.  We
discuss the numerical method in \S\ref{sec:numerics}.

Together, the flame width and the fire-polishing length set strict
limits on the densities and size of domains that can be addressed
through fully resolved simulations.  Steady state laminar flame
simulations show that we need about 5--10 zones inside the flames
thermal width,
\begin{equation}
\label{eq:thermalwidth}
l_f \equiv \frac{T_{\mathrm{ash}} - T_{\mathrm{fuel}}}{\max\{\nabla T\}} \enskip .
\end{equation}
We note that this width measurement is a factor of 2--3 narrower than
the alternate measure defined as the width of the region where the
temperature is 10\% above the fuel temperature to 90\% of the ash
temperature that was used in \citet{timmeswoosley1992}.  In order for
the RT instability to develop, we need the computational domain to be
at least one fire-polishing length wide, but in practice, we
want 10 or more fire-polishing lengths in the domain for bubble merger
to take place and the RT instability to become well developed.  These
restrictions limit us to consider densities $\le 1.5\times 10^7~\gcc$
in the present study.  The lower limit of densities we can consider is
set only by what is attained in the supernova explosion itself---we
stop at $6.67\times 10^6~\gcc$ here.

When the RT instability dominates the burning, it is the fluid motion
mixing the fuel with the hot ash that controls the burning of the new
fuel, rather than thermal conduction.  Thus, the flame has a very
large thermal width, governed by the mixing, at the lowest densities.
To initiate a detonation, the temperature in a sufficiently large
region needs to rise in unison, such that the over-pressure generated
by the burning is strong enough to drive a shock just ahead of the
burning layer.  The critical size of this region is called the
matchhead size.  If this RT thermal width becomes as large as the
critical detonation matchhead size, as tabulated in
\citet{niemeyerwoosley1997}, it is possible that a DDT can occur.
Furthermore, the size of the matchhead increases dramatically as the
density decreases, and at low densities, it may be larger than the
entire star.  We note that recent simulations
\citep{gamezo2003,reinecke2002b} indicate that a pure deflagration can
be successful in unbinding the star.  We will explore the possible of
a DDT, using the scaling behavior we find for flames in the
distributed burning regime.

\section{NUMERICAL METHODS}
\label{sec:numerics}

The simulations were performed using an adaptive, low Mach number
hydrodynamics code, as described in \citet{SNeCodePaper}.  The state
variables in the inviscid Navier-Stokes equations are expanded in
powers of Mach number, following \citet{MajSet85}.  The result is that
the pressure is decomposed into a dynamic and thermodynamic component,
the ratio of which is of order Mach number squared.  Only the dynamic
component appears in the momentum equation.  These equations are
solved using an approximate projection formalism
\citep{AlmBelColHowWel98,daybell00}, breaking the time evolution into
an advection and a projection step.  The advection is computed using
an unsplit, second-order Godunov method that updates the species and
enthalpy to the new time level and finds provisional velocities.  A
divergence constraint on the velocities is provided by the
thermodynamics, and enforced in the projection step, where the
provisional velocities are projected onto the space of vectors
satisfying this constraint.  Burning is handled through operator
splitting.  The low Mach number formulation allows us to follow the
evolution of these slow moving flames---in this density range, the
laminar speeds are $\mathrm{M} \sim 10^{-5}$---without the constraint
of the sound speed on the timestep.  This affords us timesteps that
are $O(1/\mathrm{M})$ larger than a fully compressible code would
take.  We are still free however to have large density jumps as we
cross the flame.  This method was validated for astrophysical
applications by comparing solutions of 1-d laminar flames to fully
compressible results \citep{SNeCodePaper}, and in our study of the
Landau-Darrieus instability (paper~I), where the growth rate computed
from our calculations matched the theoretical predictions across a
range of wave numbers.  The code is unchanged from the description in
\citet{SNeCodePaper}.  We review the input physics used in the present
simulations below.

This method contrasts sharply with that used in the only previous
study of the RT instability in Type~Ia SNe, as presented in
\citet{khokhlov1995}.  There, the fully compressible PPM algorithm was
used, along with a model for a thickened flame to represent it on the
small scales.  The laminar flame speed matched the correct physical
value, but the flame thickness was much larger than the physical
value.  In the present case, resolving the thermal structure frees us
from the need for a flame model.

The equation of state consists of a Helmholtz free-energy tabular
component for the degenerate\slash relativistic electrons\slash
positrons, an ideal gas component for the ions, and a blackbody
component for radiation, as described in \citet{timmes_swesty:2000}.
The conductivities contain contributions from electron-electron and
electron-ion processes, and are described in
\citet{timmes_he_flames:2000}.  All of the flames are half
carbon\slash half oxygen, but only the carbon is burned.  A single
reaction, $^{12}$C($^{12}$C,$\gamma$)$^{24}$Mg, is followed, using the
unscreened rate from \citet{caughlan-fowler:1988}.  Since we do not
expect the burning to proceed up to the iron-group elements at these
low densities, this reaction alone is sufficient to model the nuclear
reactions.  The reaction rate for oxygen burning is several orders of
magnitude slower than the carbon rate at the ash temperatures we
reach, so neglecting oxygen burning is valid on our scales.

The calculations are initiated by mapping a one-dimensional steady
state laminar flame (computed with the same code) onto our grid,
shifting the zero point according to a random phase, 10 frequency
sinusoidal perturbation.  Beginning with a steady-state flame in
pressure equilibrium virtually eliminates any transients from
disturbing the flow ahead of the flame.  In all of the results
presented, the flame is moving downward in our domain.  The transverse
boundaries are periodic, the upper boundary is outflow, and the lower
boundary is inflow, with the inflow velocity set to the laminar flame
speed.  This keeps an unperturbed flame stationary on our grid.  The
solutions presented here were computed with an advective CFL number
of~$0.225$ and typically required 8000~timesteps.  For all the runs we
present, adaptive mesh refinement was used, with a base grid plus one
finer level that has twice the resolution.  This allows us to focus
the resolution on the interesting parts of the flow, allowing larger
domains to be studied.  Refinement triggered on the temperature
gradient and the gradient of carbon mass fraction, ensuring that the
flame's reaction zone was always at the finest resolution.

\section{RESULTS}
\label{sec:results}

Table~\ref{table:simparams} lists the parameters used for all the
simulations we present.  For the lowest density case, $6.67\times
10^6~\gcc$, we actually present results for three different domain
sizes and two resolutions and for the $10^7~\gcc$ case, we present an
additional simulation with the burning disabled.  Except where noted
otherwise, all discussion of the $6.67\times 10^6~\gcc$ results refer
to the widest domain run.  In the discussion below, we first
concentrate on the change in character of the burning as the density
is changed.  Next we look at integral quantities that help
characterize the flame, followed by a measurement of the RT growth
rate and a comparison of the RT instability with and without burning.
Finally, we consider resolution and domain size effects on the results
we present.

\subsection{Varying the Density and the Transition to Distributed Burning}

\begin{figure*}
\begin{center}
\epsscale{0.6}
\plotone{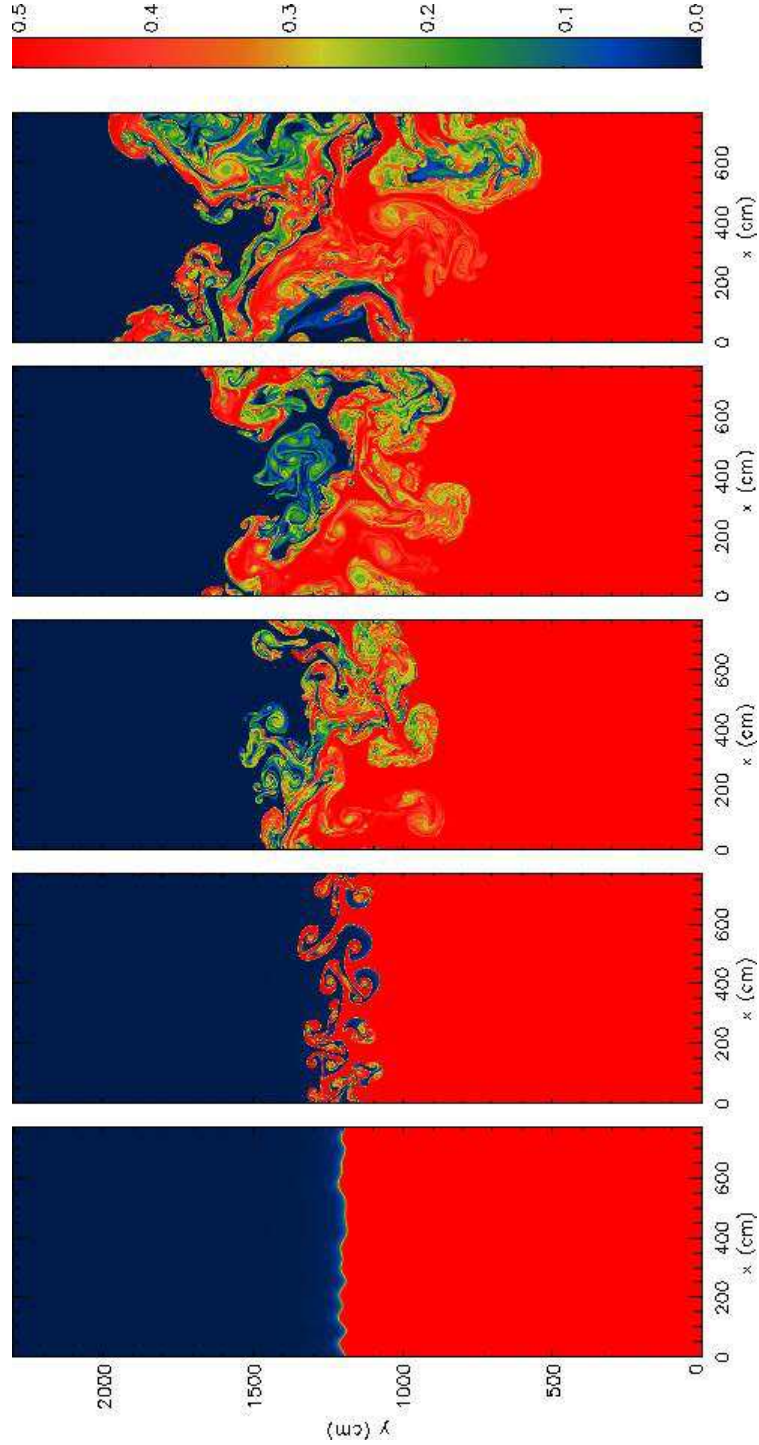}
\epsscale{1.0}
\end{center}
\caption{\label{fig:rt_6.67e6_wide} Carbon mass fraction for a 768~cm
wide, $6.67\times 10^6~\gcc$ C/O flame shown every $1.6\times
10^{-3}$~s from $0$~s until $6.4\times 10^{-3}$~s.  The fuel appears
red (carbon mass fraction = 0.5), and gravity points toward
increasing~$y$. At this low density, the RT instability dominates the
burning, and a large mixed region develops.  The flame surface is not
well defined here, a defining characteristic of the distributed
burning regime.}
\end{figure*}

\begin{figure*}
\begin{center}
\epsscale{0.7}
\vskip -0.25in
\plotone{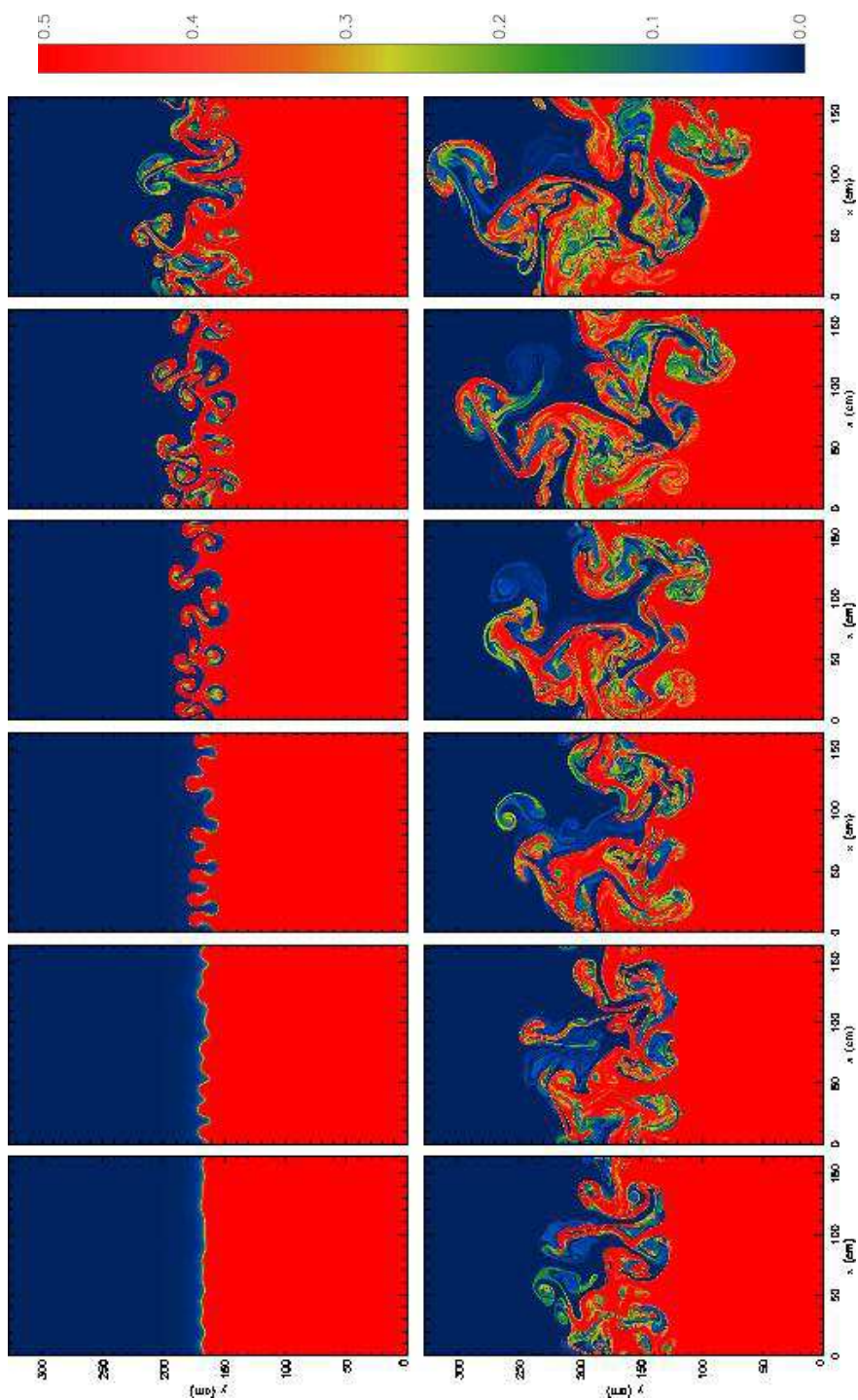}
\end{center}
\caption{\label{fig:rt_1.e7} Carbon mass fraction for $10^7~\gcc$ C/O
flame shown every $2.55\times 10^{-4}$~s from 0~s until $2.80\times
10^{-3}$~s.  The fuel appears red (carbon mass fraction = 0.5), and
gravity points toward increasing~$y$.  As the instability evolves,
fingers of denser carbon move into regions of hot ash, and gradually,
these fingers burn away, turning yellow and finally blue.  The burning
and RT growth rate are much better balanced at this density, in
comparison to the $6.67\times 10^6~\gcc$ case shown in
Figure~\ref{fig:rt_6.67e6_wide}.}
\end{figure*}

\begin{figure*}
\begin{center}
\epsscale{0.7}
\vskip -0.25in
\plotone{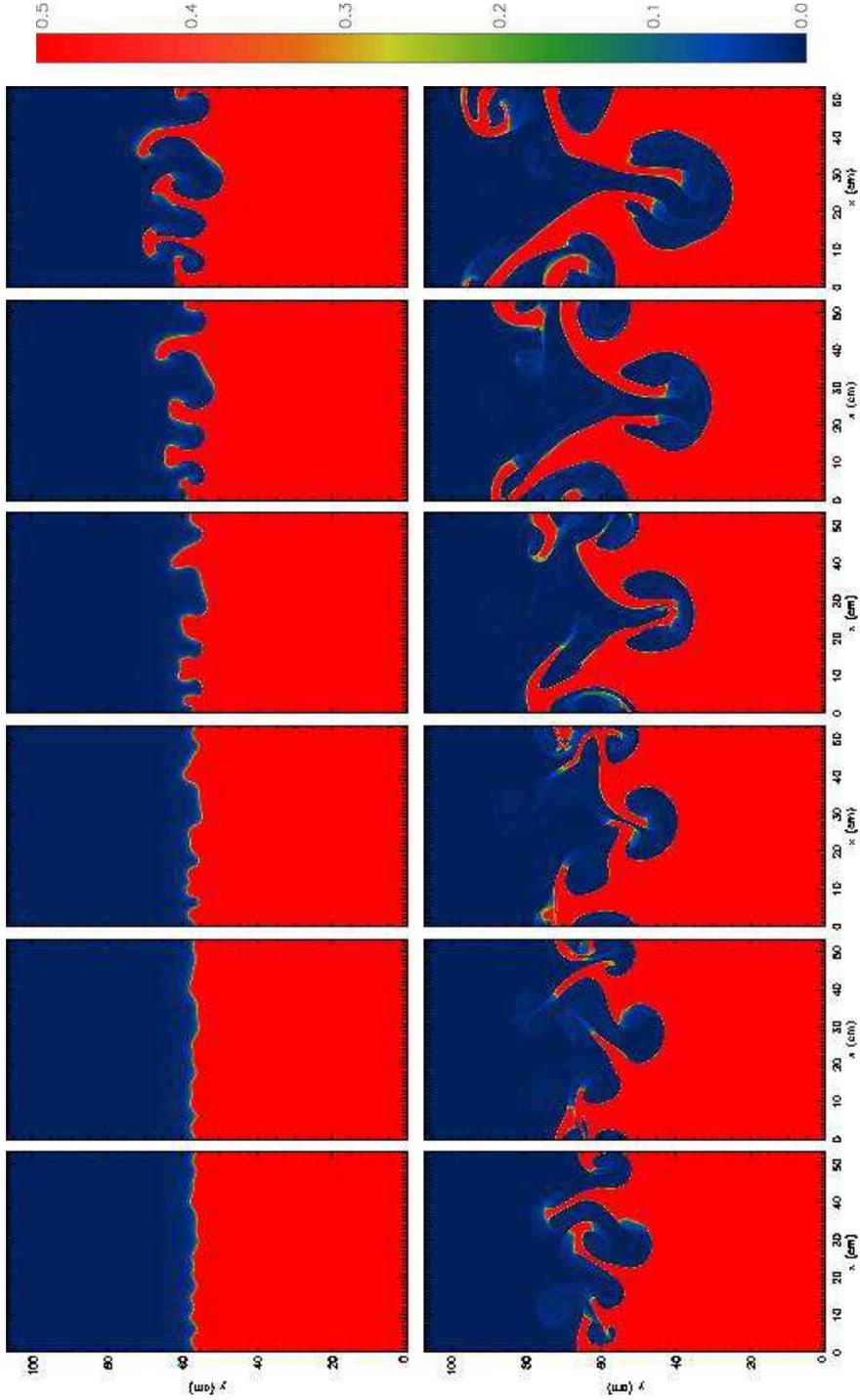}
\epsscale{1.0}
\end{center}
\caption{\label{fig:rt_1.5e7} Carbon mass fraction for $1.5\times
10^7~\gcc$ C/O flame shown every $1.33\times 10^{-4}$~s, from 0~s
until $1.46\times 10^{-3}$~s.  The fuel appears red (carbon mass
fraction = 0.5) and gravity points toward increasing~$y$.  Here, the
burning has a strong effect in moderating the RT instability.  In all
panels, it is easy to define where the flame front is---there is
always a sharp transition from the fuel to ash, a signature of the
flamelet regime.  This contrasts with the lower density runs, where
there is significant mixing.}
\end{figure*}

Figures~\ref{fig:rt_6.67e6_wide} to~\ref{fig:rt_1.5e7} show the
results for the three densities we consider in this paper, $6.67\times
10^6~\gcc$, $10^7~\gcc$, and $1.5\times 10^7~\gcc$.  The fuel is at
the bottom of the domain (and colored red), and the ash is at the top
of the domain.  Gravity points up, toward increasing~$y$, in all
simulations.  For all runs, we set $g = 10^9~\mathrm{cm~s^{-2}}$.
This value is appropriate for the outer regions of the star, and
assumes that some pre-expansion of the white dwarf has taken place.
The domain width was chosen based on the flame width and
fire-polishing length (and considerations of how many zones we can
reasonably compute) to contain several unstable wavelengths.  We see
from the figures that as the density increases, the small scale
structure is greatly diminished, as the burning is more effective in
containing it.  The calculations were run until the instability
reached the top or bottom of the computational domain, or the size of
the mixed region grew to be much larger than the width of the domain,
leading to saturation of the instability which will be discussed
later.  In the descriptions below, `bubble' refers to the hot ash
buoyantly rising into the cool fuel, and `spike' refers to the cool
fuel falling into the hot ash.

The results for $\rho = 6.67\times 10^6~\gcc$, 768~cm wide simulation
(Figure~\ref{fig:rt_6.67e6_wide}) shows the flow quickly becoming
mixed and the initial perturbations rapidly forgotten.  A mixed region
500--1000~cm wide forms, separating the fuel and ash.  Some smaller
features begin to burn away as the flame evolves and the ash gets
entrained by the fluid motions.  At late times, it becomes impossible
to draw a well defined flame surface separating the fuel and ash, as
the RT mixing clearly dominates.  In \S\ref{sec:res}, we look at the
effect of different domain widths on this growth.

The $10^7~\gcc$ run (Figure~\ref{fig:rt_1.e7}) shows a balance between
the RT growth and the burning.  Spikes of fuel push well into the hot
ash, forming well defined mushroom caps.  Before these caps can travel
too far into the ash, they burn away.  This is seen for several spikes
in Figure~\ref{fig:rt_1.e7}---the mushrooms gradually burn away,
almost uniformly, turning yellow\slash green, and finally leaving a
ephemeral outline in the ash as the burning completes.  As the
instability evolves, we see longer wavelength modes beginning to
dominate, but they also begin to burn away.  We are prevented from
watching the final dominant mode that grows burn away because of it
reaching the top of our domain.

The highest density case, $1.5\times 10^7~\gcc$
(Figure~\ref{fig:rt_1.5e7}) behaves qualitatively different than the
two lower densities.  A defining characteristic of this density is
that the flame front is always sharply defined---a trait that we
expect to carry over to higher densities as well.  At this density,
the burning proceeds rapidly enough that it can suppress the RT
instability significantly.  This is as expected, since the
fire-polishing length is larger than the flame width at this density
(see Table~\ref{table:flameproperties}).  As the spikes of fuel extend
into the hot ash, we see them rapidly burning away from the outside
inward.  The RT mushroom caps hardly have time to form before their
ends are burned away, giving them a hammerhead-like appearance.

\begin{figure*}
\begin{center}
\plotone{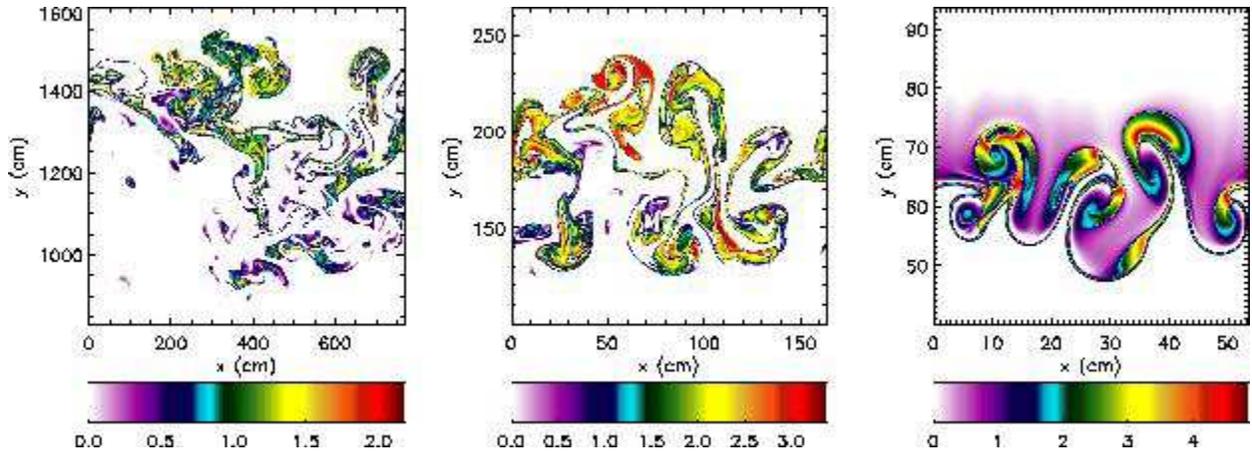}
\end{center}
\caption{\label{fig:carbon_destruction} Log of carbon destruction
rate, $|\dot{\omega_C}|$, for the three
densities, $6.67\times 10^6~\gcc$ (left), $10^7~\gcc$ (center), and
$1.5\times 10^7~\gcc$ (right).  At the lowest density, the peak
burning regions are distributed throughout the reactive region in
several `hot' spots.  The reactive region becomes more laminar at
higher densities, where the burning is more effective at curtailing
the RT instability.}
\end{figure*}

Figure~\ref{fig:carbon_destruction} shows the carbon destruction rate,
$|\dot{\omega_C}|$, at the midpoint of each calculation for the three
densities.  At the low density end, the burning is concentrated in
small regions whereas at the high end, the burning is smoother.  We
would expect this trend to continue for densities outside the range we
consider here.  At higher densities, the burning will dominate even
more, and we would expect the flame surface to be well defined on the
small scales.  At densities lower than $6.67\times 10^6~\gcc$, the
mixed region will grow even larger, as the burning becomes even less
effective at suppressing the RT instability.  We will look at how the
width of the reactive region grows in proportion to that of the mixed
region at these low densities in \S\ref{sec:growth}.

\subsection{\label{sec:wrinkle} Wrinkling and Acceleration of the Flame}

Having described the overall evolution of the flames, we now present
some quantitative measures of flame behavior.  We define the effective
flame speed, $V_\mathrm{eff}$, in terms of the carbon consumption in
the domain; namely,
\begin{equation}
V_\mathrm{eff}(t) = -\frac{\int_\Omega \rho \dot\omega_C d{\bf x}}
             {W (\rho X_C)^\mathrm{in}}
\end{equation}
where $\Omega$ is the spatial domain of the burning region, $W$ is the
width of inflow face, $(\rho X_C)^\mathrm{in}$ is the inflow carbon
mass fraction, and $\rho\dot\omega_C$ is the rate of consumption of
carbon due to nuclear burning.  Accurate direct evaluation of this
integral is problematic because of the operator split treatment of
reactions in the numerical method.  A robust and accurate estimate of
$V_\mathrm{eff}$ can be obtain by integrating the conservation
equation for carbon mass fraction $\partial(\rho X_C)/\partial t +
\nabla\cdot(\rho u X_C) = -\rho\dot\omega_C$ over $\Omega$ and a time
interval $[T_1,T_2]$ to obtain
\begin{equation}
V_\mathrm{eff} = \frac{\int_\Omega(\rho X_C)|^{T_2}_{T_1}d{\bf x}}{(T_2-T_1) W (\rho X_C)^\mathrm{in}}-u^\mathrm{in}
\end{equation}
$u^\mathrm{in}$ is the inflow velocity.  Note that this formula is
only valid when essentially all carbon is being consumed in the
domain.

Another quantity of interest is the area of the flame, which for our
two-dimensional studies is a length. Computing a flame length for the
lower density cases is problematic since there is often not a distinct
flame surface. In this paper we define a flame length as the number of
zone edges where the carbon mass fraction passes through~0.25.  We
normalize the length to the initial flame length so we can measure the
growth of the burning surface.  This approach, while crude, is robust
and defines a reasonable method for computing the length of a well
mixed flame where other approaches are infeasible.  In particular,
this approach systematically overestimates the flame length; however,
this can be compensated for by normalizing by the initial flame
length.  We have validated that for well-defined flames, this method
agrees well with the length of a contour as computed with the
commercial {\tt IDL} package, after normalization.

\begin{figure*}[H]
\begin{center}
\epsscale{0.6}
\plotone{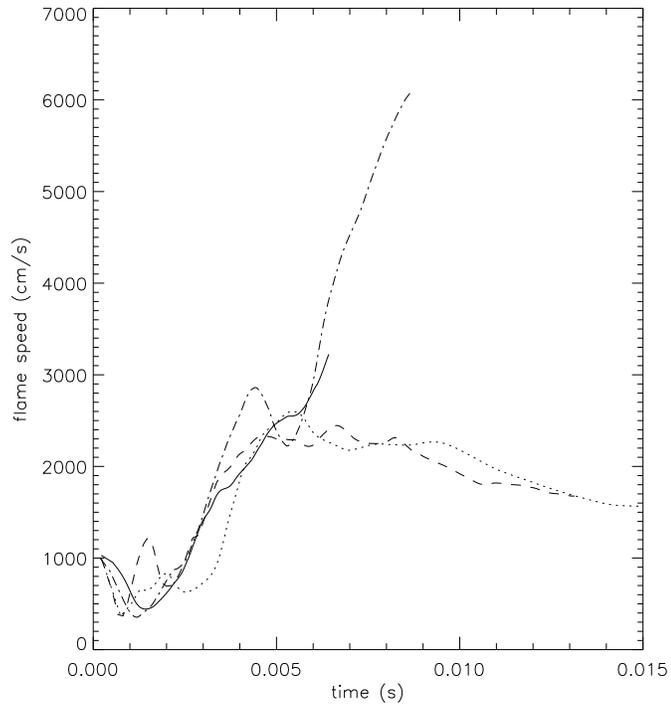}
\end{center}
\caption{\label{fig:rt_6.67e6_speeds} Flame speed as a function of
time for the $6.67\times 10^6~\gcc$ C/O flame RT simulations.  The
solid line is the main simulation at this density (768~cm wide
domain).  Also shown are the results from three narrower runs, 384~cm
(dot-dash), and 96~cm at high resolution (dashed) and low resolution
(dotted)---see Table~\ref{table:simparams}.  The peak and asymptotic
values of the velocity in the 96~cm domain seem to be insensitive to
the resolution, demonstrating convergence.  The wider domain allows
for longer wavelength modes to go unstable resulting in a larger
velocity, as expected.}
\end{figure*}

\begin{figure*}[H]
\begin{center}
\epsscale{0.6}
\plotone{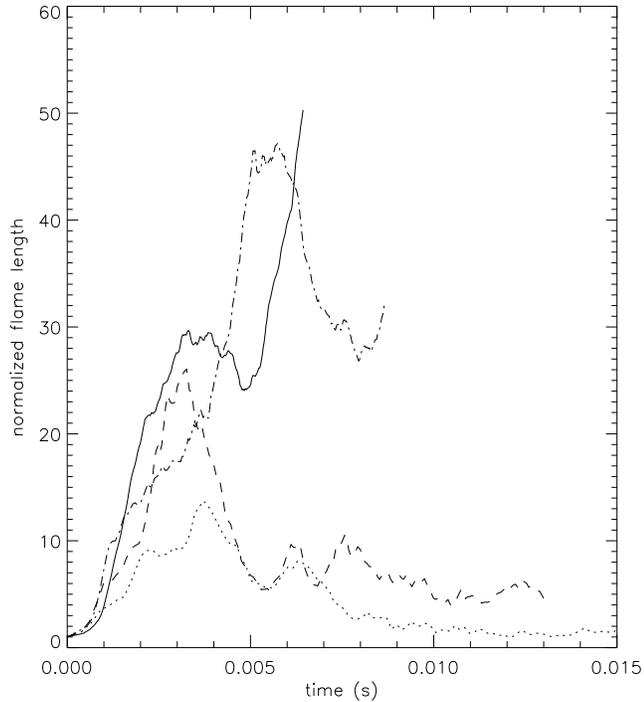}
\end{center}
\caption{\label{fig:rt_6.67e6_length} Normalized length of the flame,
$L(t)/L(t=0)$, as a function of time for the $6.67\times 10^{6}~\gcc$
RT simulations.  The standard (768~cm wide) case is shown as the
solid line.  Also shown are three narrower runs, 384~cm wide
(dot-dash), and 96~cm wide at high resolution (dashed) and low
resolution (dotted).  The wider domain allows for longer
wavelength modes to grow, leading to the greater increase in flame
length as compared to the narrow run.  The resolution study shows the
finer resolved run producing more area, which is not completely
unexpected, given the almost fractal like character of these flames.}
\end{figure*}

\begin{figure*}[H]
\begin{center}
\epsscale{0.6}
\plotone{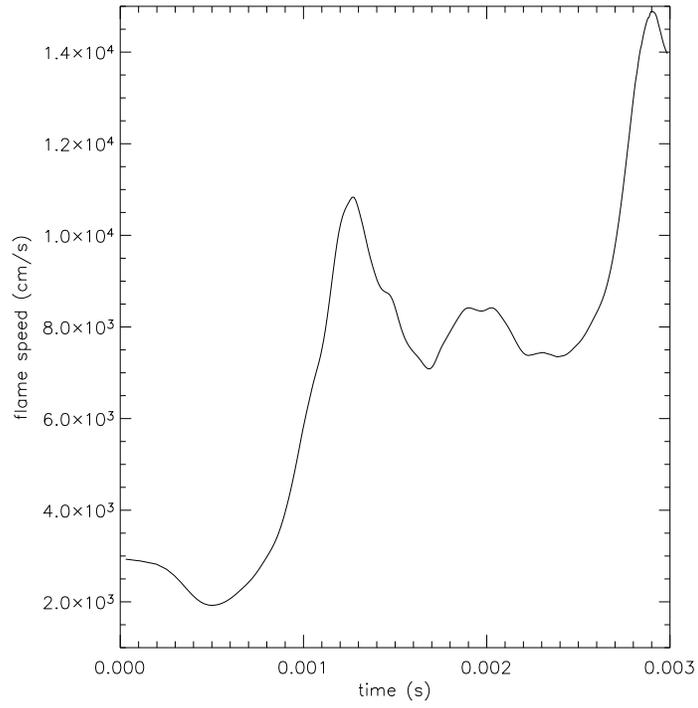}
\end{center}
\caption{\label{fig:rt_1.e7_speeds} Flame speed as a function of time
for the $10^{7}~\gcc$ RT simulation.  Considerable acceleration
(factor of $5$) is seen as the instability evolves.  This run shows
no signs of saturating, but is stopped because the spikes of fuel
reach the top of the domain.}
\end{figure*}

\begin{figure*}[H]
\begin{center}
\epsscale{0.6}
\plotone{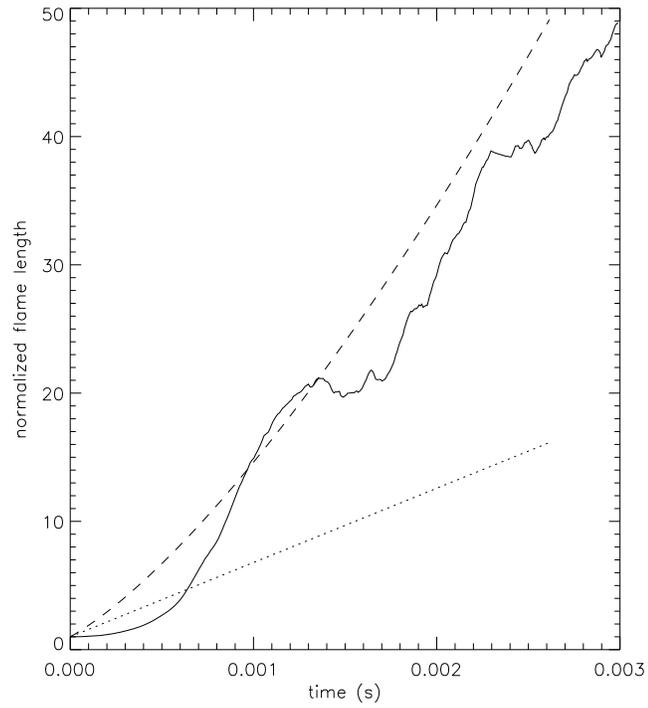}
\end{center}
\caption{\label{fig:rt_1.e7_length} Normalized length of the flame,
$L(t)/L(t=0)$, as a function of time for the $10^{7}~\gcc$ RT
simulation.  Through the course of the simulation, the surface of the
flame increases to 50 times its original length.  The dotted and
dashed lines are predictions from the fractal model for surface growth
(Eq.~[\ref{eq:fractal}]), using $D=1.5$ and $D=1.7$ respectively.
Here, the $D=1.7$ curve provides a reasonable fit to the numerical
data.}
\end{figure*}

\begin{figure*}[H]
\begin{center}
\epsscale{0.6}
\plotone{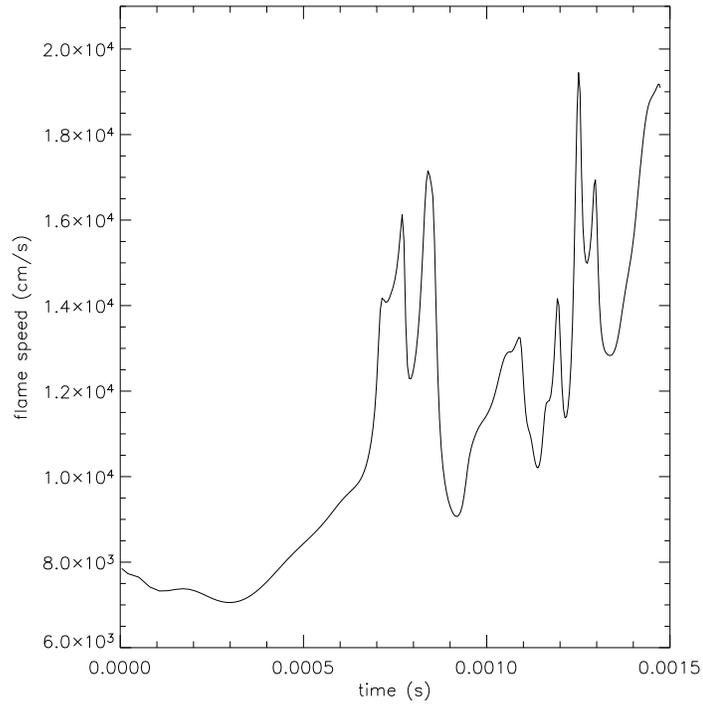}
\end{center}
\caption{\label{fig:rt_1.5e7_speeds} Flame speed as a function of time
for the $1.5\times 10^{7}~\gcc$ RT simulation.  At this density,
the velocity is very choppy---reaching a peak and then rapidly falling
away as a plume of fuel is consumed, but the overall trend is of
increasing velocity, leading to a speedup of about 2.5 times the laminar 
speed.}
\end{figure*}

\begin{figure*}[H]
\begin{center}
\epsscale{0.6}
\plotone{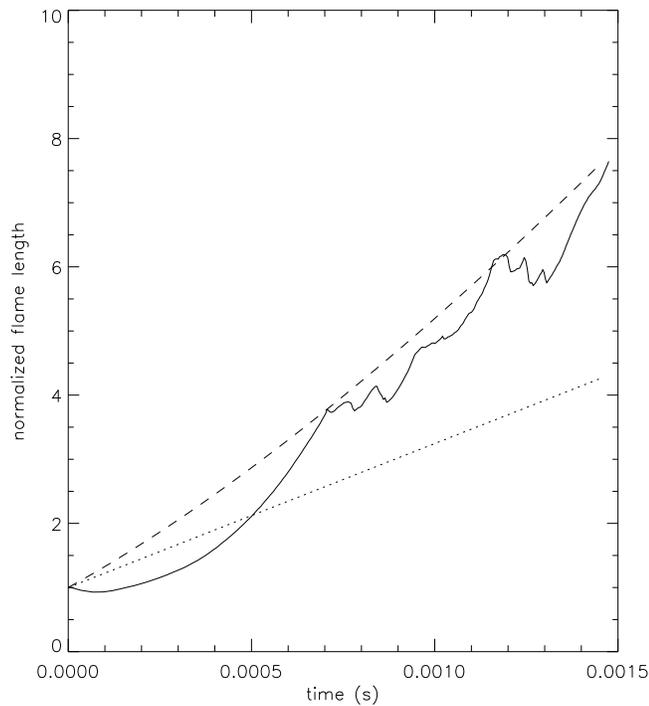}
\end{center}
\caption{\label{fig:rt_1.5e7_length} Normalized length of the flame,
$L(t)/L(t=0)$, as a function of time for the $1.5\times 10^{7}~\gcc$
RT simulation.  At this density, the growth of the flame length is
much smaller than the lower densities (see
Figures~\ref{fig:rt_6.67e6_length} and \ref{fig:rt_1.e7_length}),
reflecting the increased influence of the burning on the RT
instability.  The dotted and dashed lines are predictions from the
fractal model for surface growth (Eq.~[\ref{eq:fractal}]), using
$D=1.5$ and $D=1.7$ respectively.  As in the $10^7~\gcc$ case, the
$D=1.7$ curve provides a good fit.}
\end{figure*}

Figures~\ref{fig:rt_6.67e6_speeds} to~\ref{fig:rt_1.5e7_length} show
the speeds and lengths of the flames at the three densities as a
function of time (the additional curves on the $6.67\times 10^6~\gcc$
plots will be discussed in \S \ref{sec:res}).  The maximum speedups we
observe are 3~times the laminar speed for the $6.67\times 10^6~\gcc$
run, 5~times for the $10^7~\gcc$ run, and 2.5~times for the $1.5\times
10^7~\gcc$ run.  These speedups are much larger than any seen in our
companion study on the Landau-Darrieus instability (paper~I), as
expected, based on the outcome of the large scale simulations of the
explosion.  The flame length increases by a factor of 50 throughout
the simulation for the two lowest densities, but only increases to by
a factor of 8 for the $1.5\times 10^7~\gcc$ case, due to the strong
moderation of the RT instability by the burning.

\begin{figure*}[H]
\begin{center}
\epsscale{0.6}
\plotone{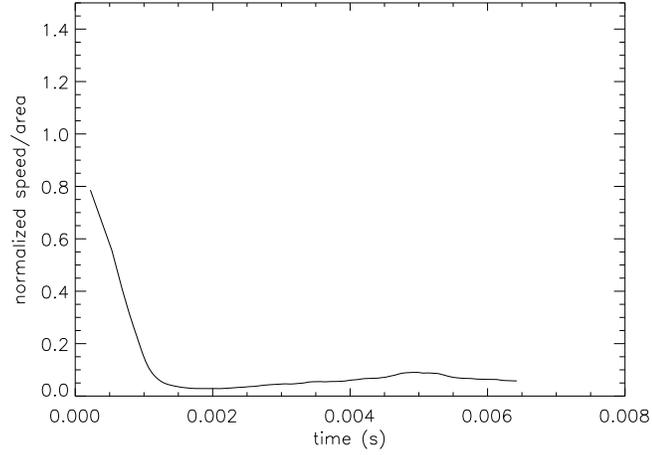}
\end{center}
\caption{\label{fig:rt_6.67e6_v_vs_a} Normalized speed vs.\ area,
$[v(t)/v_\mathrm{laminar}]/[l(t)/W]$, as a function of time for
$6.67\times 10^6~\gcc$ RT unstable flame in the 768~cm wide domain.
Here we see that the speedup is significantly less than the
geometrical prediction from the growth in flame surface area.}
\end{figure*}

\begin{figure*}[H]
\begin{center}
\epsscale{0.6}
\plotone{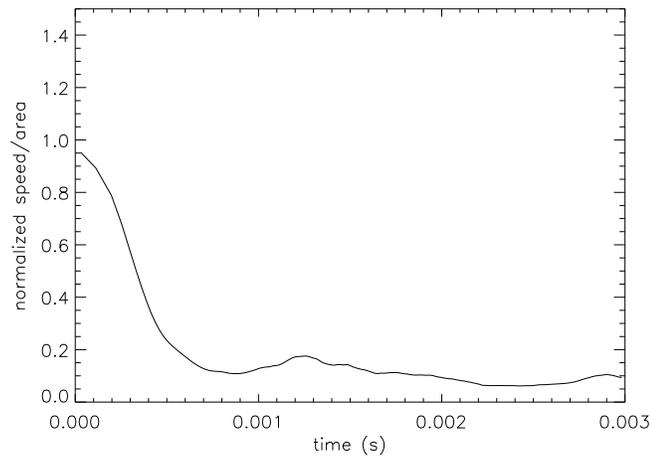}
\end{center}
\caption{\label{fig:rt_1.e7_v_vs_a} Normalized speed vs.\ area,
$[v(t)/v_\mathrm{laminar}]/[l(t)/W]$, as a function of time for
$10^7~\gcc$ RT unstable flame.  Again, we see significant departure
from the geometrical prediction for the growth in flame surface area.}
\end{figure*}

\begin{figure*}[H]
\begin{center}
\epsscale{0.6}
\plotone{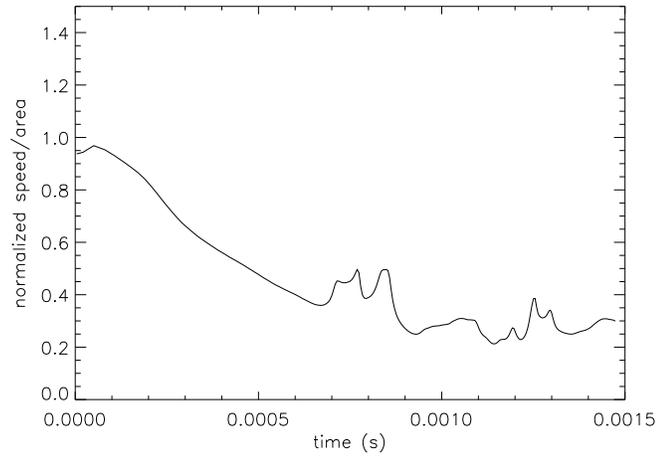}
\end{center}
\caption{\label{fig:rt_1.5e7_v_vs_a} Normalized speed vs.\ area,
$[v(t)/v_\mathrm{laminar}]/[l(t)/W]$, as a function of time for
$1.5\times 10^7~\gcc$ RT unstable flame.  This case is the closest yet
to achieving the simple geometrical scaling for the flame speed, but
still falls short, asmyptoting to about 30\% of the geometrical
prediction.}
\end{figure*}

Comparing the velocity curves to the length curves, we see that, in
general, the velocity is not strictly proportional to the flame
surface area.  This is not unexpected, since the stretch and curvature
that the flame experiences as well as the complicated motions of the
RT instability pushing the fuel and ash together will modify the local
burning rate. Evidence of this variability is seen in
Figure~\ref{fig:carbon_destruction} where considerable variation in the
carbon destruction rate is observed, even for the highest density
case.  To further quantify this effect, in
Figures~\ref{fig:rt_6.67e6_v_vs_a} to~\ref{fig:rt_1.5e7_v_vs_a} we
plot the velocity divided by flame surface area normalized to
$v_\mathrm{laminar}/W$, where $W$ is the width of the box.  With the
caveat that there are ambiguities in defining flame surface area for
the ``distributed'' flames, we observe that, after an initial
transient, the flame speed versus area relaxes to an essentially
statistically stationary value.  For the highest density case, the
effective flame velocity is approximately 30\% of the laminar flame
speed $\times$ the area enhancement. For the lower density cases, the
value drops to approximately 10\%.  Thus, at these densities, we find
\begin{equation}
\label{eq:v_vs_a}
v(t) = c ~v_{\mathrm{laminar}} \frac{A(t)}{A_0} \enskip ,
\end{equation}
where $c$ is a proportionality constant not equal to 1.  This
proportionality constant increases with density, reflecting the fact
that the fire-polishing wavelength, which sets the smallest scale on
which we can wrinkle the flame, grows as well.  Thus at higher
densities, the localized curvature is smaller, and as a result, the
localized velocity is less affected.  This has strong implications for
subgrid models, particularly in the flamelet regime, where it is
normally assumed that the velocity scales directly as the increase in
area of the flame surface.  It appears that this is not the case, and
is further complicated by the density dependence of the
proportionality constant.

\cite{woosley1990} introduced a fractal model to describe the growth
of the flame surface in a supernova when the burning is in the
flamelet regime.  The self-similar range through which a fractal
description applies is bounded by the fire-polishing length on the
small scales and the Sharp-Wheeler description for the RT instability
growth on the large scales.  In two-dimensions, a fractal model for
the growth of the flame surface would be
\begin{equation}
\label{eq:fractal}
L = L_0 \left ( \frac{\lambda_{\mathrm{max}}}{\lambda_{\mathrm{min}}} \right )^{D-1} = L_0 \left(\frac{\alpha g_{\mathrm{eff}}^2 (t-t_0)^2}{4\pi v_{\mathrm{laminar}}^2} \right )^{D-1} \enskip ,
\end{equation}
where we expect $1 < D < 2$ in the self-similar scaling regime.  The
offset in time, $t_0$, is chosen such that $L = L_0$ at $t=0$.  The
value of $\alpha$ is taken to be $0.05$.  As we will see in \S
\ref{sec:growth}, the Sharp-Wheeler model may not be the proper
description of the reactive RT instability.  With that caveat,
Figures~\ref{fig:rt_1.e7_length} and \ref{fig:rt_1.5e7_length}, we
plot the predictions of the fractal scaling model using $D=1.5$ and
$1.7$, which give area growing as $t$ and $t^{1.4}$ respectively.
From these figures, it seems that, if this fractal model is correct,
the growth in the flame surface is closely represented by a fractal
dimension of 1.7.  This agrees with the value computed from
two-dimensional RT studies by \citet{hasegawa1996}.  Some evidence
suggests that the fractal dimension of RT induced turbulence may
increase with time \citep{dimotakis1998}.  We caution however that we
only span about a decade between $\lambda_{\mathrm{min}}$ and
$\lambda_{\mathrm{max}}$ in these simulations.  If such a scaling
holds, it could serve as a subgrid model for large scale simulations,
however, the evolution of the proportionality constant in
Equation~(\ref{eq:v_vs_a}) needs to be understood.  Additional
calculations that encompass a larger range of spatial scales are
needed to further validate this model.

\begin{figure*}[H]
\begin{center}
\epsscale{0.5}
\plotone{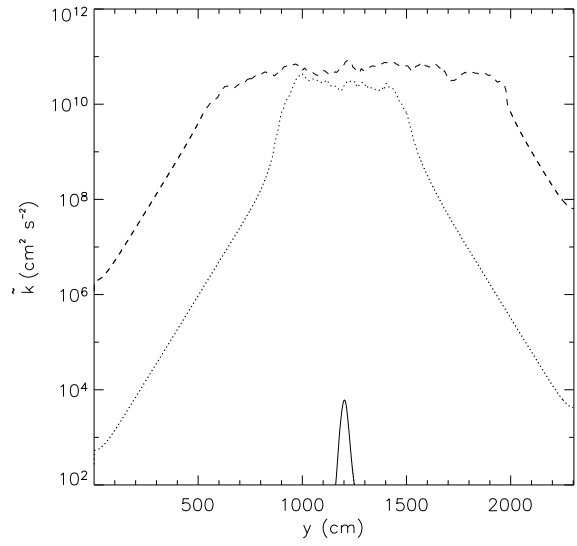}
\end{center}
\caption{\label{fig:rt_6.67e6_tke_y} Turbulent kinetic energy,
$\tilde{k}$, as a function of height for the 768~cm wide $6.67\times
10^6~\gcc$ RT unstable flame three times: $0$~s (solid), $3.2\times
10^{-3}$~s (dot), and $6.4\times 10^{-3}$~s (dash).  As the RT flame
evolves, the turbulent kinetic energy monotonically increases, with
the value roughly constant within the mixed region.}
\end{figure*}

\begin{figure*}[H]
\begin{center}
\epsscale{0.5}
\plotone{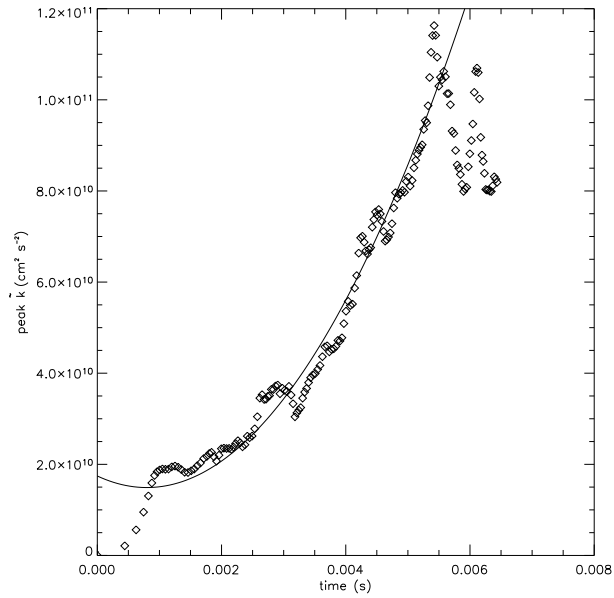}
\end{center}
\caption{\label{fig:rt_6.67e6_tke_peak} Peak turbulent kinetic energy,
$\tilde{k}$, as a function of time for the 768~cm wide $6.67\times
10^6~\gcc$ RT unstable flame.  The solid line is a quadratic fit to
the first $0.0055$~s of evolution.  At late times, the growth
stagnates, as the mixed region becomes larger than the width of the
box.}
\end{figure*}

\begin{figure*}[H]
\begin{center}
\epsscale{0.5}
\plotone{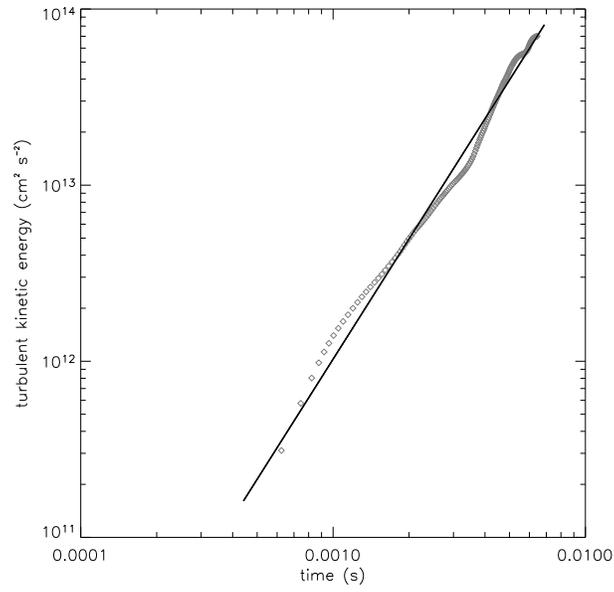}
\end{center}
\caption{\label{fig:rt_6.67e6_tke} Total turbulent kinetic energy as a
function of time, $k(t)$, for the 768~cm wide $6.67\times 10^6~\gcc$ RT
unstable flame.  The symbols are the data and the solid line is a fit,
$k(t) = 6.514\times 10^{18} t^{2.267}$.}
\end{figure*}

\begin{figure*}[H]
\begin{center}
\epsscale{0.5}
\plotone{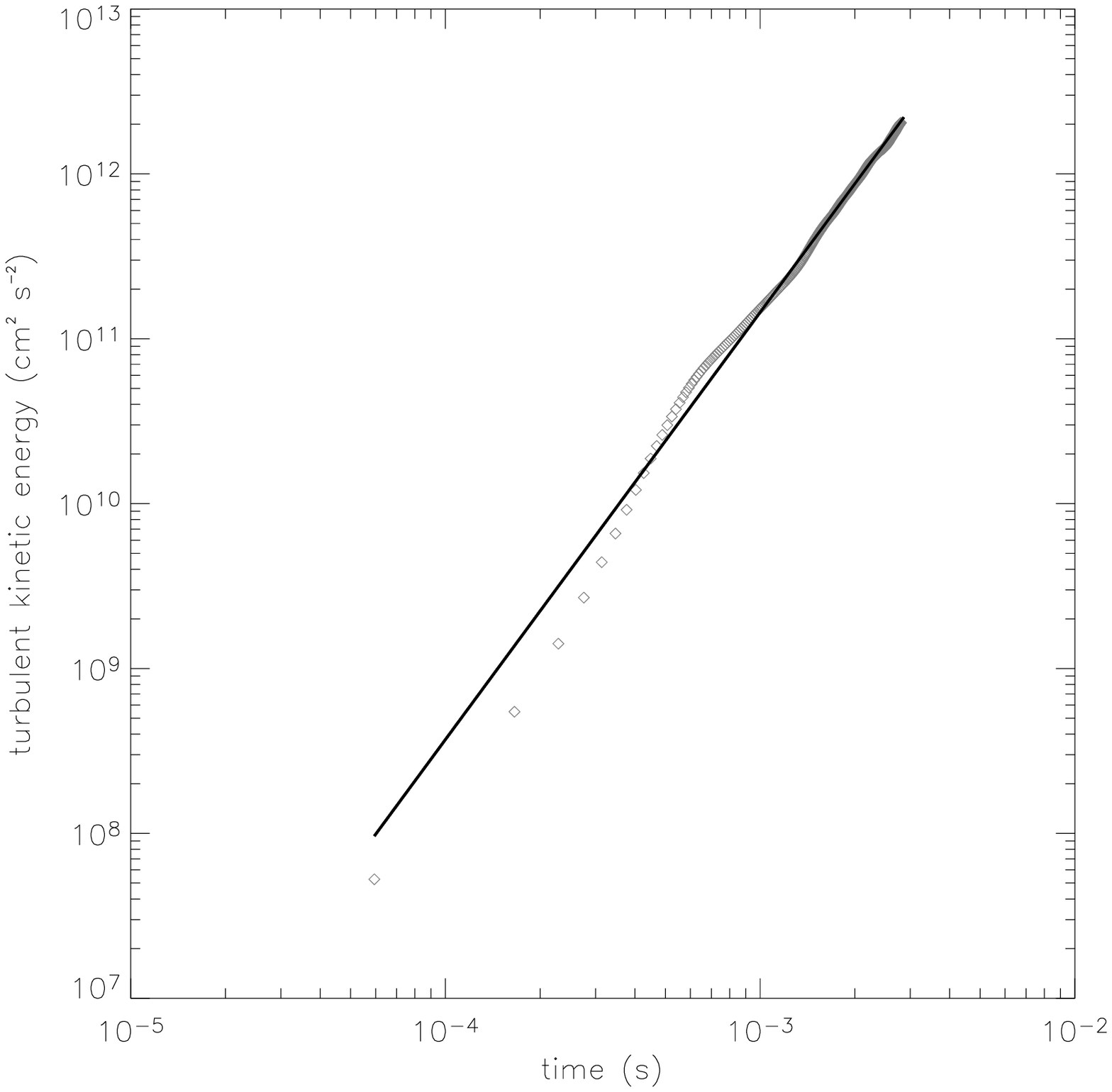}
\end{center}
\caption{\label{fig:rt_1.e7_tke} Total turbulent kinetic energy as a
function of time, $k(t)$, for the $10^7~\gcc$ RT unstable flame.  The
symbols are the data and the solid line is a fit, $k(t) = 8.597\times
10^{18} t^{2.591}$.}
\end{figure*}

\begin{figure*}[H]
\begin{center}
\epsscale{0.5}
\plotone{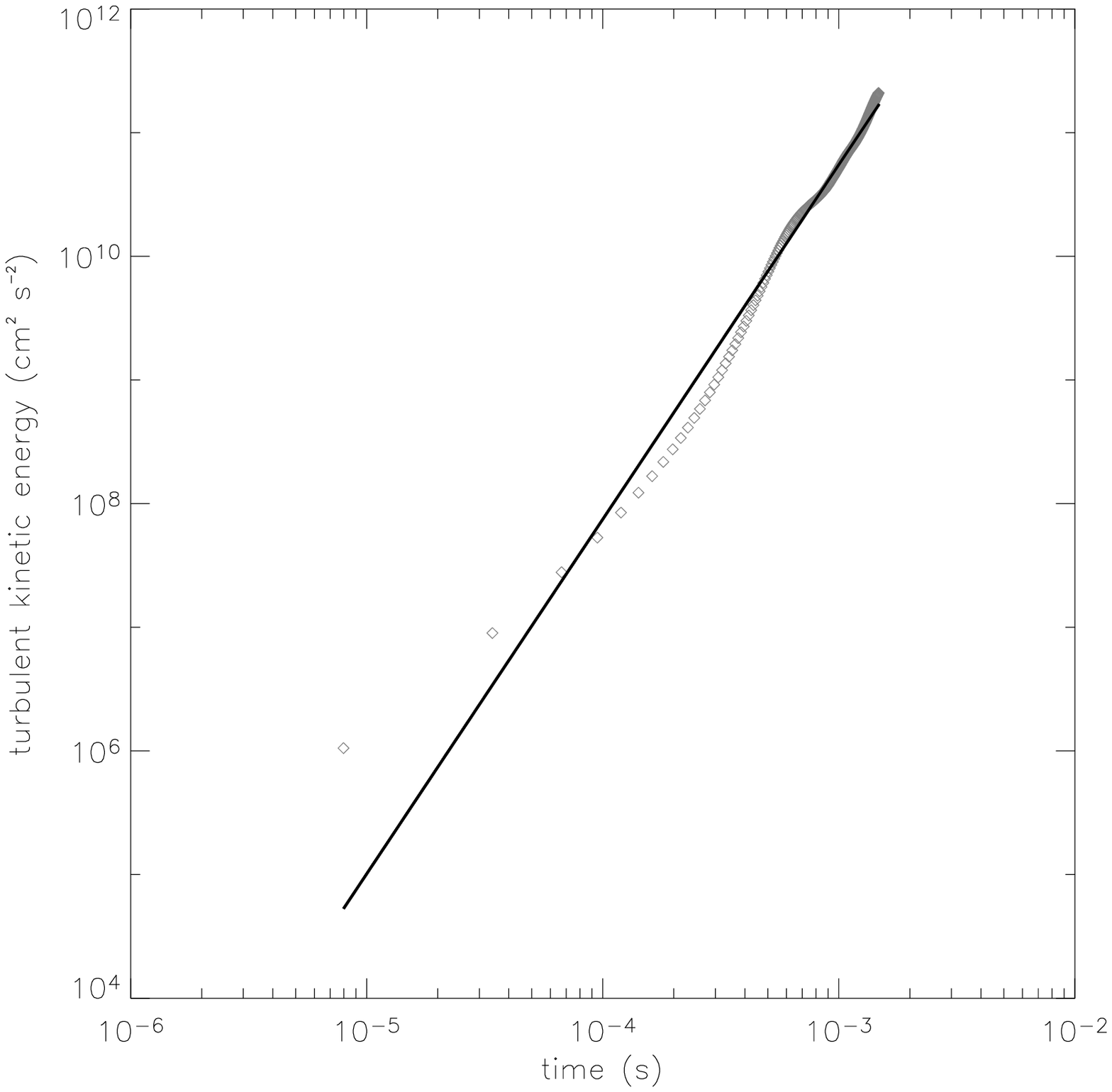}
\end{center}
\caption{\label{fig:rt_1.5e7_tke} Total turbulent kinetic energy as a
function of time, $k(t)$, for the $1.5\times 10^7~\gcc$ RT unstable
flame.  The symbols are the data and the solid line is a fit, $k(t) =
2.202\times 10^{19} t^{2.867}$.}
\end{figure*}

Not all of the energy release by the flame goes into driving the
expansion of the star.  The RT unstable flame and associated
Kelvin-Helmholtz instabilities also generate turbulence.  Turbulence
diagnostics can be difficult to define---here we use an integral
quantity, the Favre average turbulent kinetic energy which plays a
role in the $k-\epsilon$ subgrid model for turbulence.  The Favre
average turbulent kinetic energy (for a discussion of which see
\citealt{peters:2000}), can be written
\begin{equation}
\tilde{k}(y,t) = \frac{1}{2} \left<(\rho {\bf v} - \left<\rho {\bf v}\right> )^2\right>/\left<\rho\right>^2
\label{EQ:ftke}
\end{equation}
where $\left<\cdot\right>$ denotes the horizontal spatial average of
the quantity enclosed in the brackets.
Figure~\ref{fig:rt_6.67e6_tke_y} shows $\tilde{k}$ as a function of
height at several instances in time for the $6.67\times 10^6~\gcc$
flame.  We see that, within the mixed region, the turbulent kinetic
energy is roughly constant, and this plateau rises quadratically with
time, as shown in Figure~\ref{fig:rt_6.67e6_tke_peak}.  The integral
of~$\tilde{k}(y,t)$ over the vertical extent of the domain is the
total turbulent kinetic energy~$k(t)$.  The generation of turbulent
kinetic energy as a function of time is shown for the three densities
in Figures~\ref{fig:rt_6.67e6_tke} to~\ref{fig:rt_1.5e7_tke}.  The
data suggest that the increase in kinetic energy as a function of time
is well approximated by a power law of the form
\begin{equation}
k(t) = a t^b \enskip ,
\end{equation}
are shown in Figures~\ref{fig:rt_6.67e6_tke}
to~\ref{fig:rt_1.5e7_tke}, and summarized in Table~\ref{table:tke}.
The power law fits, summarized in Table~\ref{table:tke} match the data
quite well, giving an exponent, $b = 2.3$--$2.8$, depending on the
density, with the power increasing with increasing density.  In the
lowest density run, there is considerable turbulent energy, and, as
Figure~\ref{fig:rt_6.67e6_tke_y} shows, $\sim
10^{11}~\mathrm{erg~g^{-1}}$ is reached after $6.4\times 10^{-3}$~s.
This is on a scale of $O(10^3)$~cm.  \citet{niemeyerhillebrandt1995b}
presumed pre-existing turbulent kinetic energy of this level on scales
of $10^6$~cm in their main calculation.  Assuming a one-third power
scaling of the energy cascade, on their length scales the RT generate
turbulence already overwhelms this pre-existing turbulence, and it is
continuing to grow, apparently quadratically with time.  These results
suggest that the RT instability alone can provide the turbulent
motions presumed to exist in the star.


\subsection{\label{sec:growth} The Growth Rate of RT Unstable Flames}

In the absence of any reactions, the extent of the mixed region should
increase as $t^2$, as predicted by the Sharp-Wheeler model,
Equation~(\ref{eq:sw}).  This relation is only expected to hold during
the phase where bubbles are merging, which excludes the initial linear
growth range.  Also, because we are resolving the flame structure, the
interface between the fuel and ash is not infinitesimally thin, which
complicates the definition of the Atwood number.  Furthermore, at late
times, once the bubble merging in our domain has stopped and the size
of the mixed region rivals the width of the domain, this relation will
also break down.  This relation was derived for the purely
hydrodynamical RT instability.  Burning can have large consequences
here.  The Sharp-Wheeler model describes the growth of this mixed
region, which will contain fuel and ash, but that fuel is burning, and
as it turns into ash, we would expect to find the mixed region smaller
than that of the purely hydrodynamical case.  For the present
simulations, since we have a limited number of modes in our box
(typically 10), we are only going to be able to do the fits on a small
subset of the data.  A future study will focus on scaling in the
flamelet regime, using much larger domains.

\begin{figure*}
\begin{center}
\epsscale{1.0}
\plotone{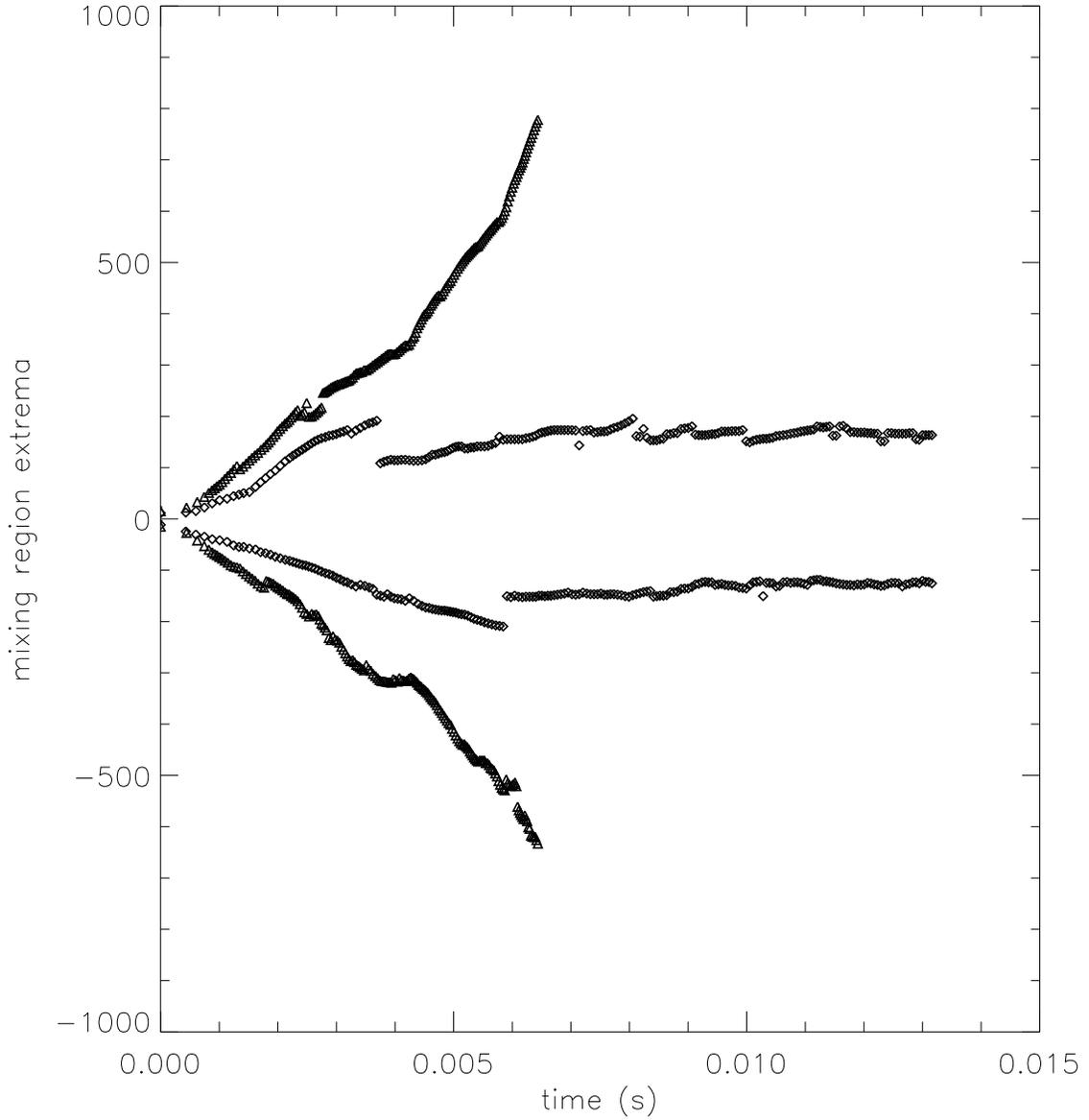}
\end{center}
\caption{\label{fig:rt_6.67e6_width} Thickness of the mixed carbon
region as a function of time for $6.67\times 10^6~\gcc$ C/O flame
RT simulation, as measured from the initial position of the
interface.  `$\Diamond$' marks the position in the laterally averaged
flame profile where the carbon mass fraction crosses 0.05 and 0.45
respectively for the 96~cm wide domain. `$\triangle$' corresponds to
the wider 768~cm domain.  We see for the narrow case, the evolution
of the flame is halted much earlier, and into a much thinner flame.}
\end{figure*}

\begin{figure*}
\begin{center}
\epsscale{1.0}
\plotone{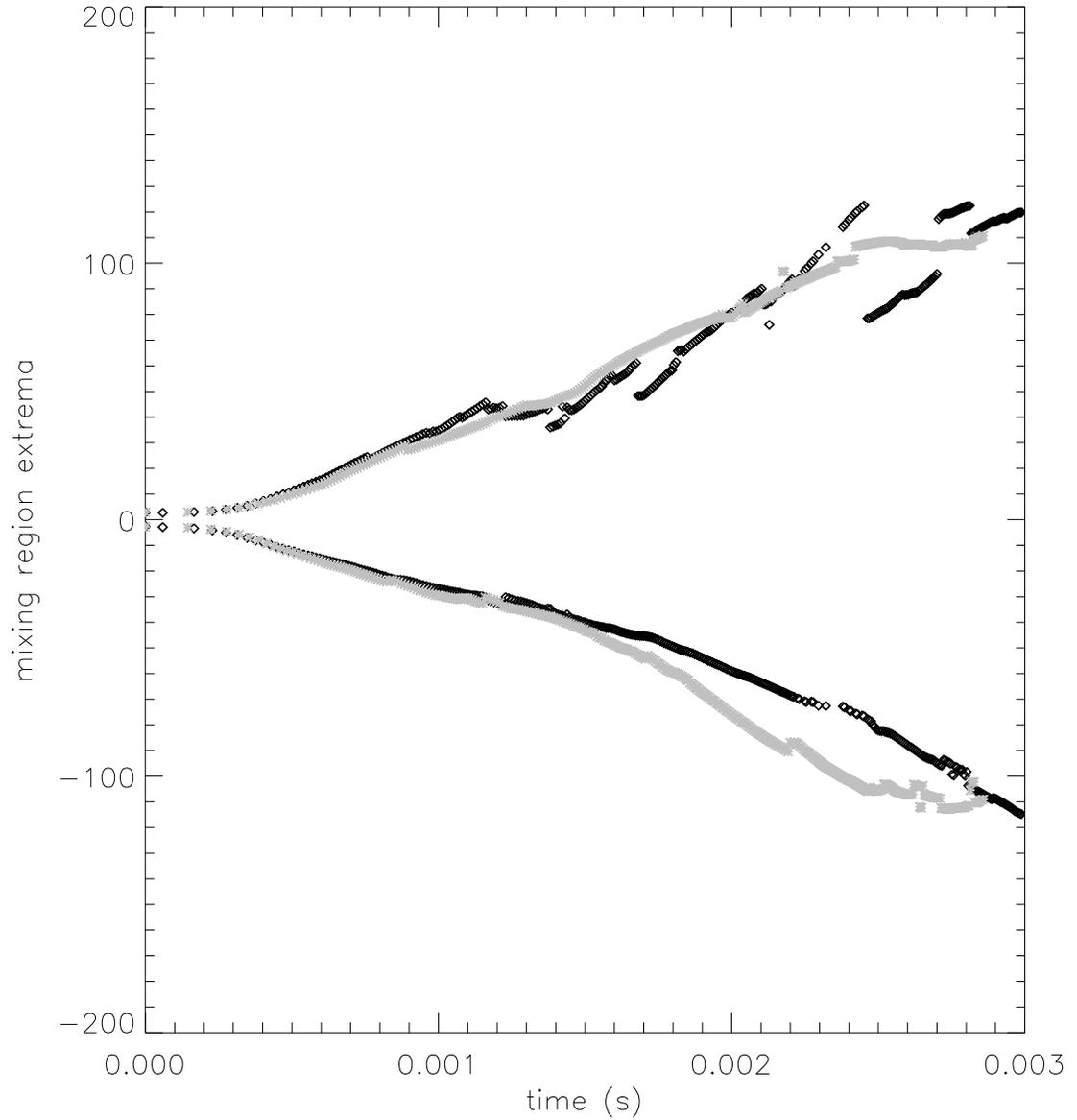}
\end{center}
\caption{\label{fig:rt_1.e7_width} Extent of the mixed region as
measured from the initial interface for both a reactive and
non-reactive $10^7~\gcc$ RT simulation as a function of time.
Symbols mark the first occurrence of carbon mass fractions of 0.05 and
0.45 in the laterally averaged profiles.  The black symbols have
burning enabled, whereas the gray symbols have no burning---all other
parameters are the same.  The effects of burning are immediately clear
in this comparison, as the `no burning' run yields considerably smoother
growth curves.}
\end{figure*}

\begin{figure*}
\begin{center}
\epsscale{1.0}
\plotone{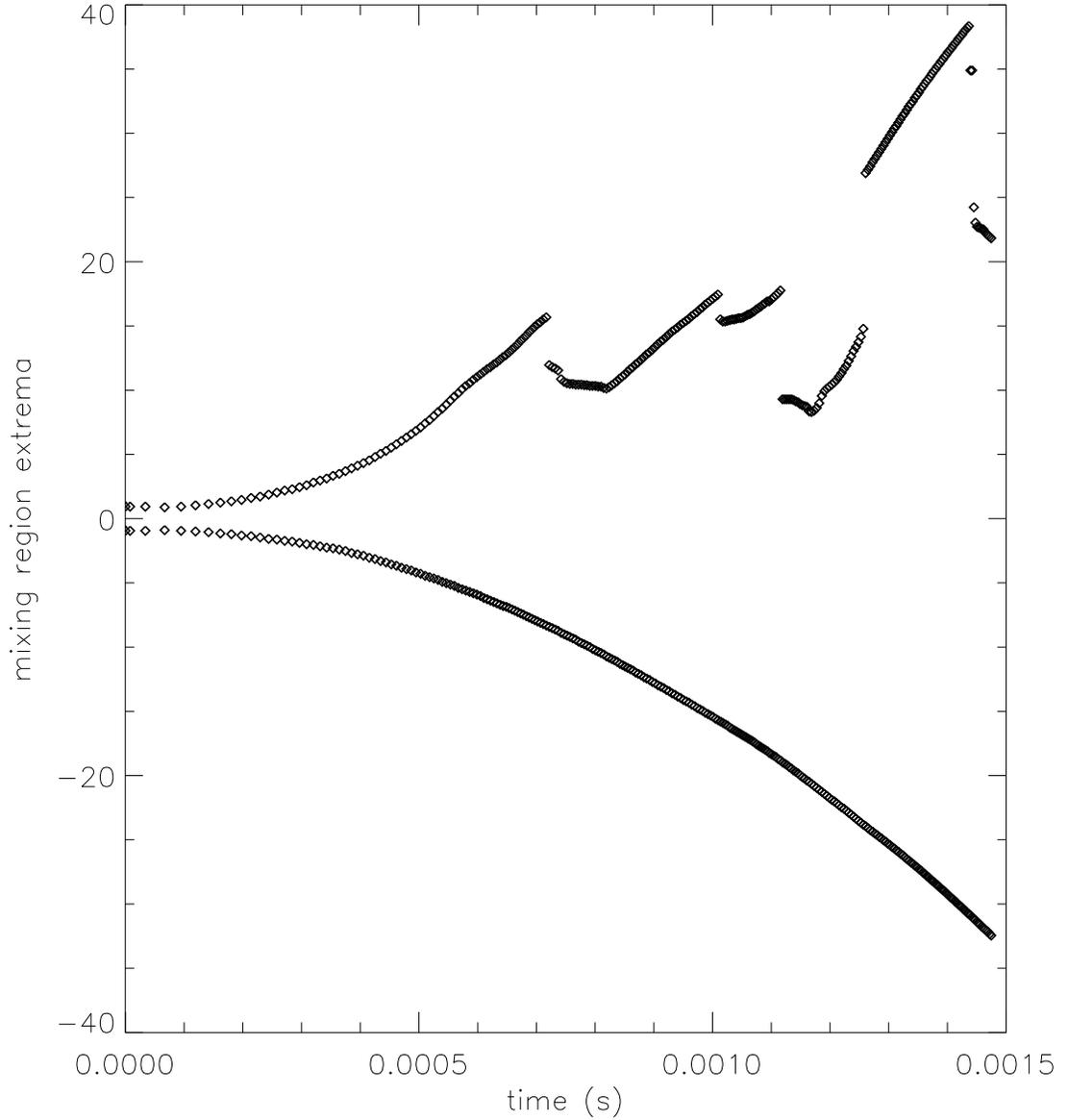}
\end{center}
\caption{\label{fig:rt_1.5e7_width} Extent of the mixed region as
measured from the initial interface for the $1.5\times 10^7~\gcc$ RT
simulation as a function of time.  Symbols mark the first occurrence
of carbon mass fractions of 0.05 and 0.45 in the laterally averaged
profiles.  The lower curve is the bubble of hot ash pushing into the
cool fuel.  Because there is nothing to burn away here, it is very
smooth, and clearly shows a $t^2$ behavior.}
\end{figure*}

Figures~\ref{fig:rt_6.67e6_width} to~\ref{fig:rt_1.5e7_width} show the
extrema of the mixed region, computed by laterally averaging the
carbon mass fractions and finding the positions where it first
exceeds~0.05 and~0.45.  This definition is consistent with that used
in \citet{khokhlov1995}.  The curves are measured with respect to the
initial position of the interface.  The top curve measures the
position the spikes of fuel pushing into the region of hot ash.  The
lower curve is the position of the bubbles of hot ash floating into
the fuel.  The sharp kinks in the spike curves represent the instances
where the plume of fuel, having pushed far into the ash region, burned
away.  At the highest density (Figure~\ref{fig:rt_1.5e7_width}), the
bubble curve is very smooth.

\begin{figure*}
\begin{center}
\epsscale{0.7}
\vskip -0.25in
\plotone{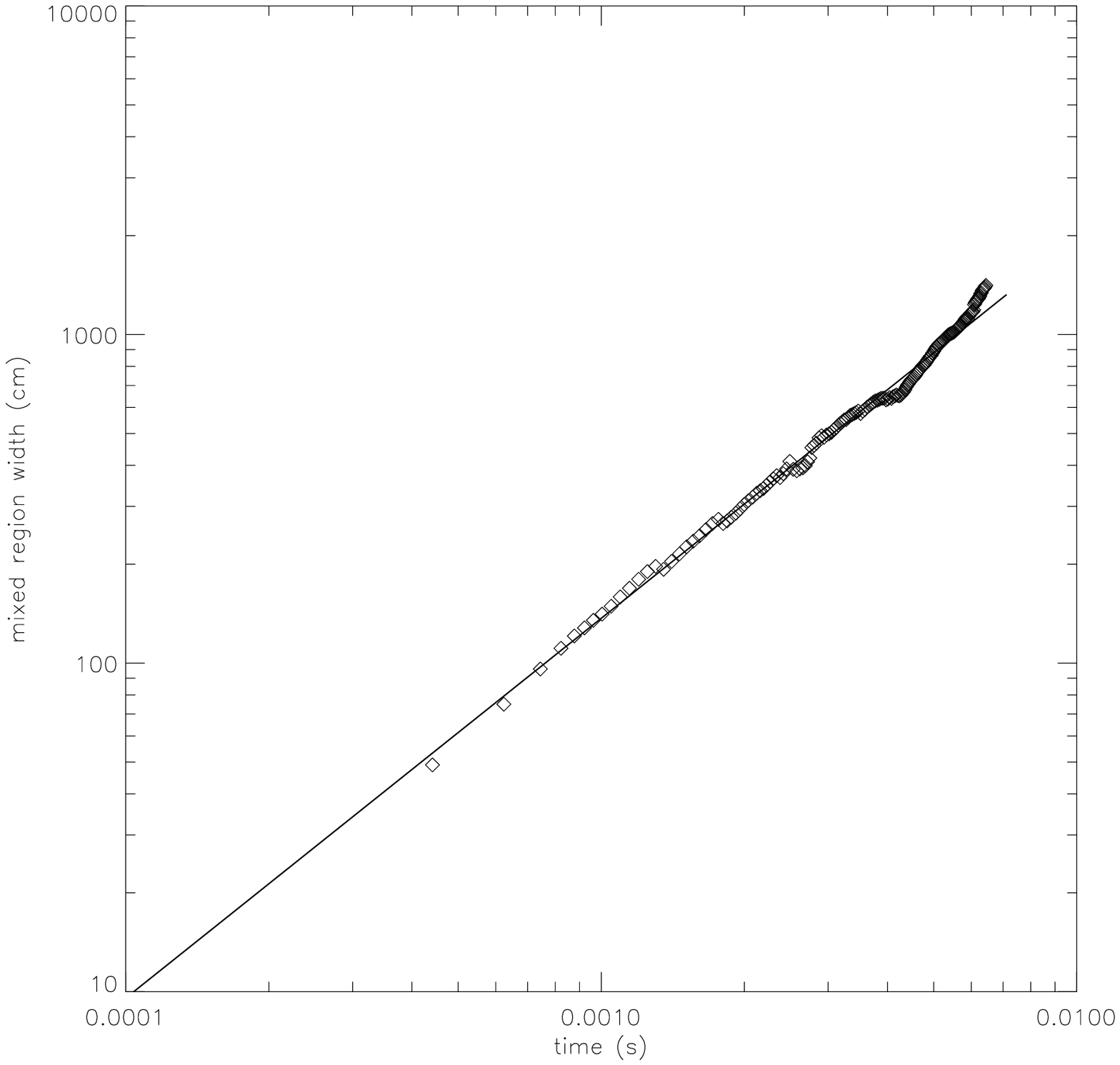}
\end{center}
\caption{\label{fig:rt_6.67e6_width_fit} Width of the mixed region for
the $6.67\times 10^6~\gcc$ RT simulation with a power law fit to time,
$w(t) = c t^n$ (solid line), with n = 1.16.  The fit was performed
over the time interval [$3\times 10^{-4}$~s, $6\times 10^{-3}$~s].}
\end{figure*}

\begin{figure*}
\begin{center}
\epsscale{0.7}
\vskip -0.25in
\plotone{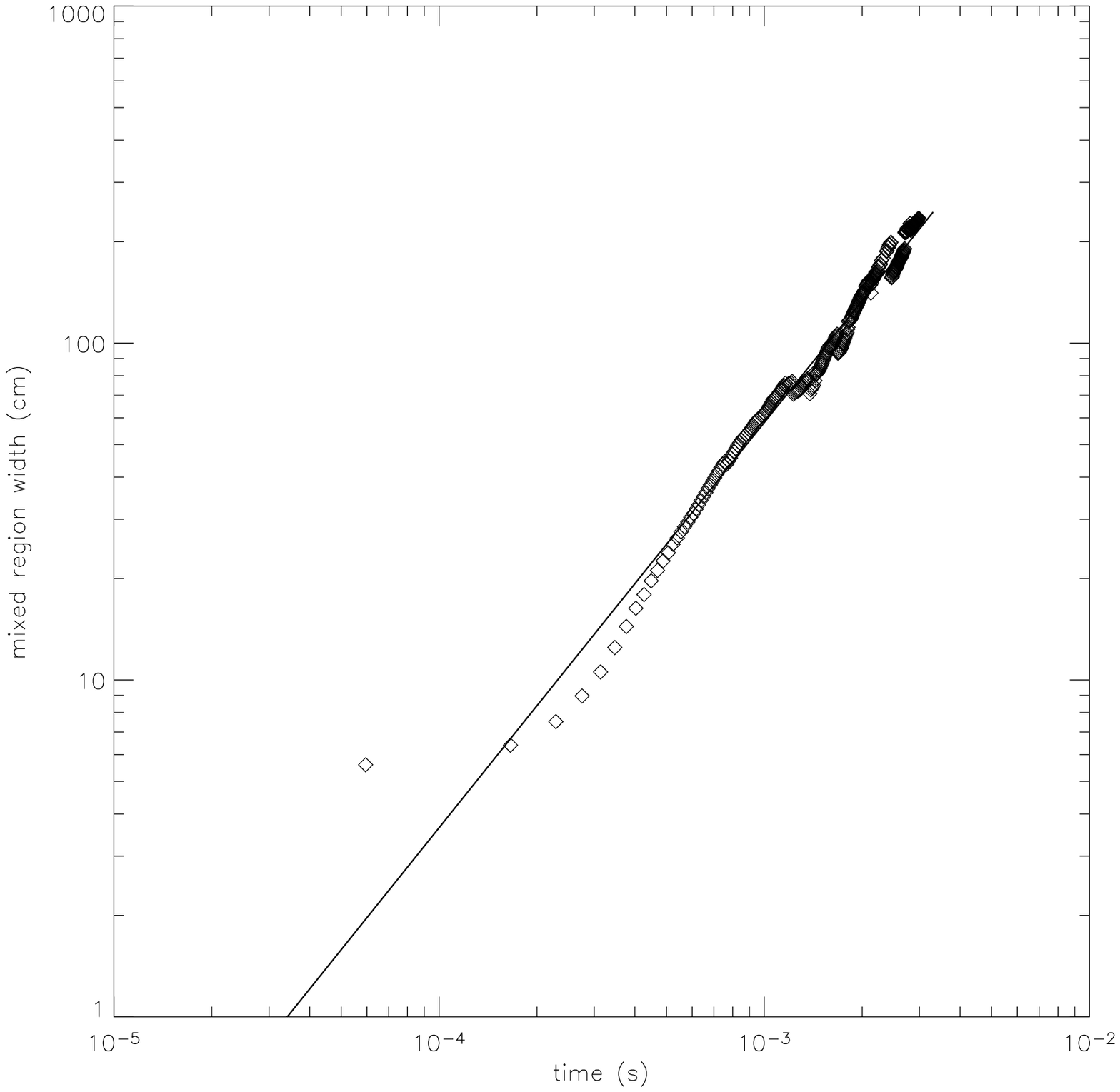}
\end{center}
\caption{\label{fig:rt_1.e7_width_fit} Width of the mixed region for
the $10^7~\gcc$ RT simulation with a power law fit to time,
$w(t) = c t^n$ (solid line), with n = 1.2.  The fit was performed
over the time interval [$5\times 10^{-4}$~s, $2.8\times 10^{-3}$~s].}
\end{figure*}

\begin{figure*}
\begin{center}
\epsscale{0.7}
\vskip -0.25in
\plotone{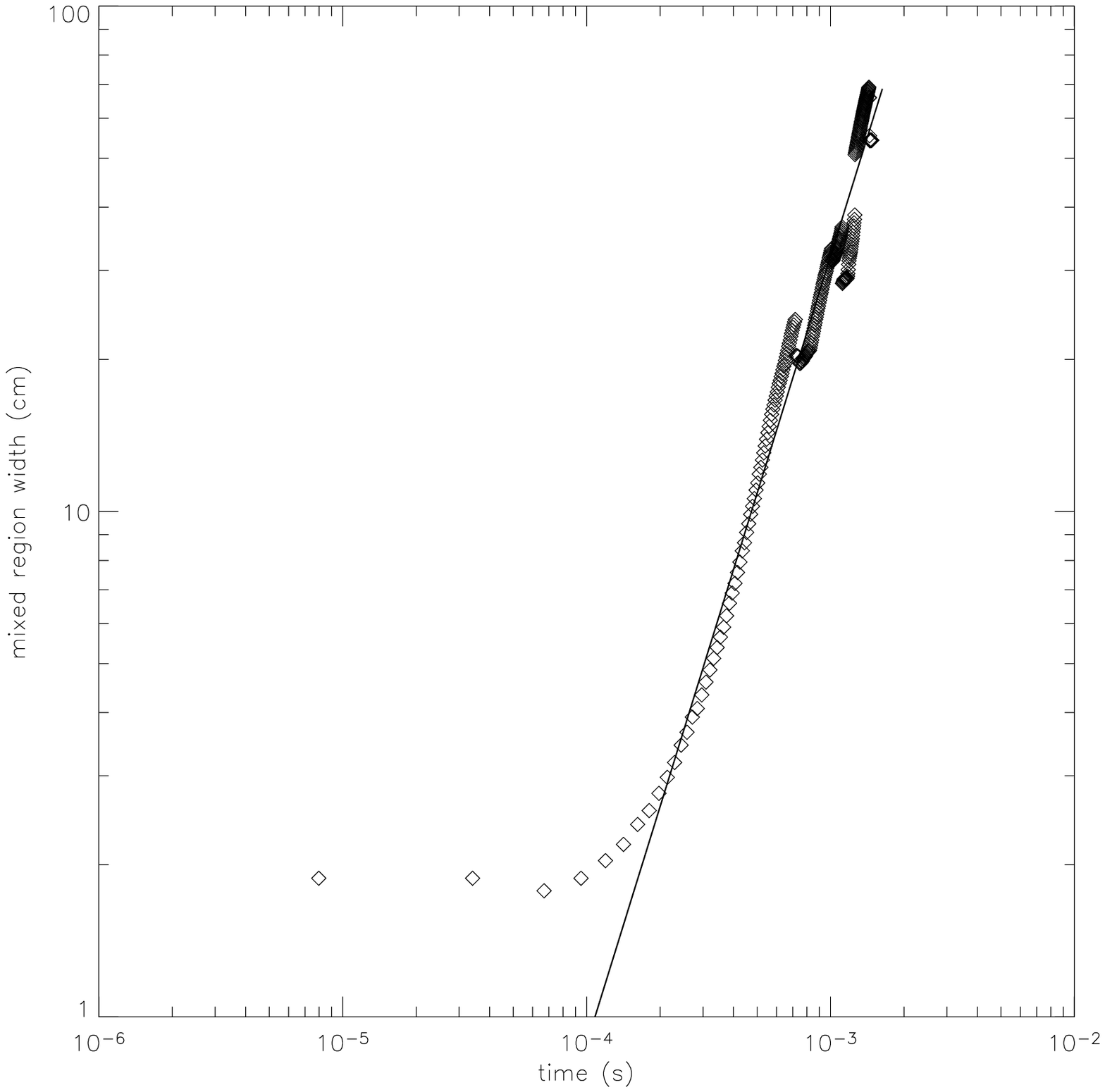}
\end{center}
\caption{\label{fig:rt_1.5e7_width_fit} Width of the mixed region for
the $1.5\times 10^7~\gcc$ RT simulation with a power law fit to time,
$w(t) = c t^n$ (solid line), with n = 1.55.  The fit was performed
over the time interval [$3\times 10^{-4}$~s, $1.45\times 10^{-3}$~s].}
\end{figure*}

\begin{figure*}
\begin{center}
\epsscale{0.7}
\vskip -0.25in
\plotone{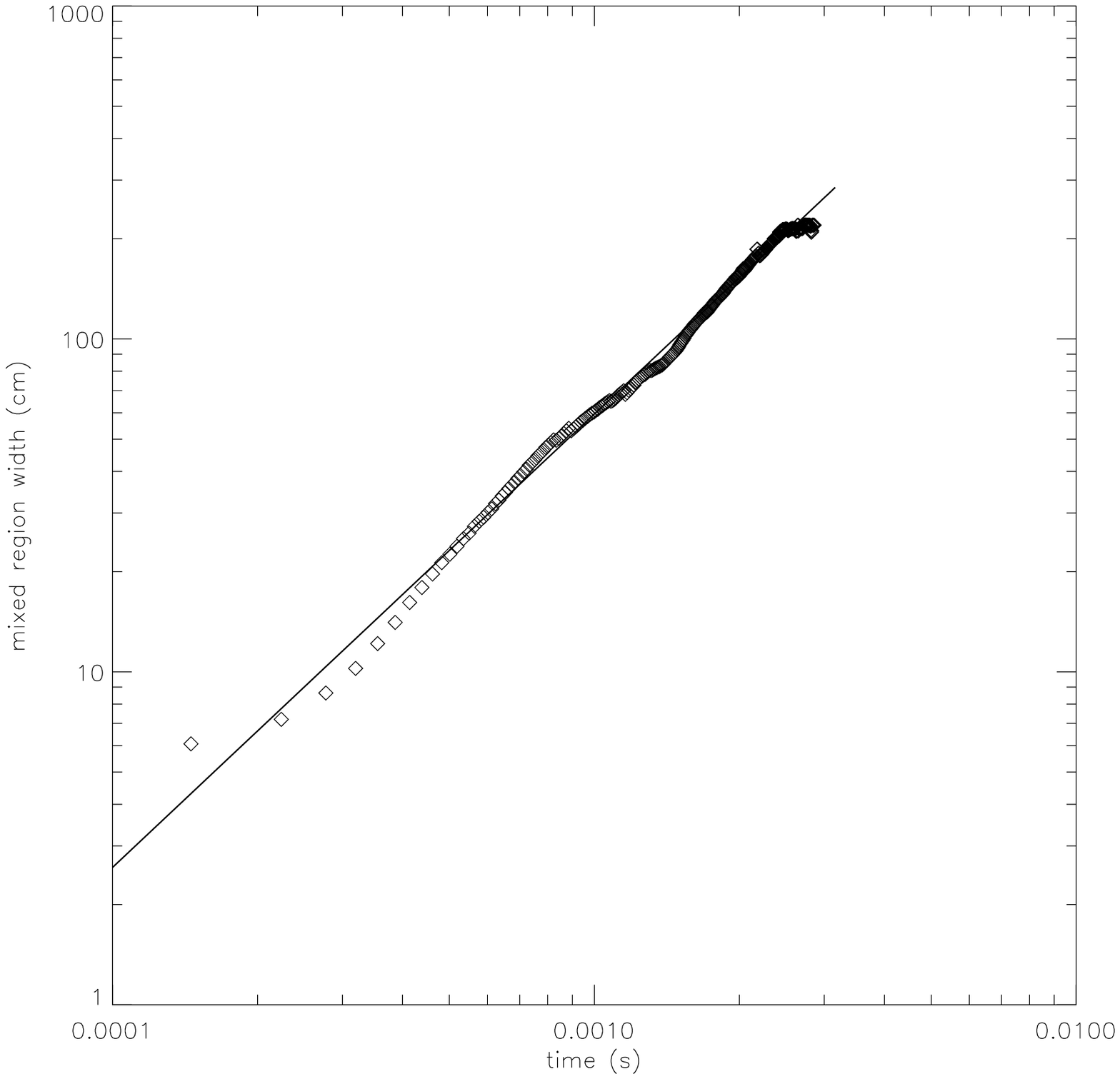}
\end{center}
\caption{\label{fig:rt_1.e7_noburn_width_fit} Width of the mixed region for
the non-reactive $10^7~\gcc$ RT simulation with a power law fit to time,
$w(t) = c t^n$ (solid line), with n = 1.36.  The fit was performed
over the time interval [$5\times 10^{-4}$~s, $2.4\times 10^{-3}$~s].}
\end{figure*}

\begin{figure*}
\begin{center}
\epsscale{0.7}
\vskip -0.25in
\plotone{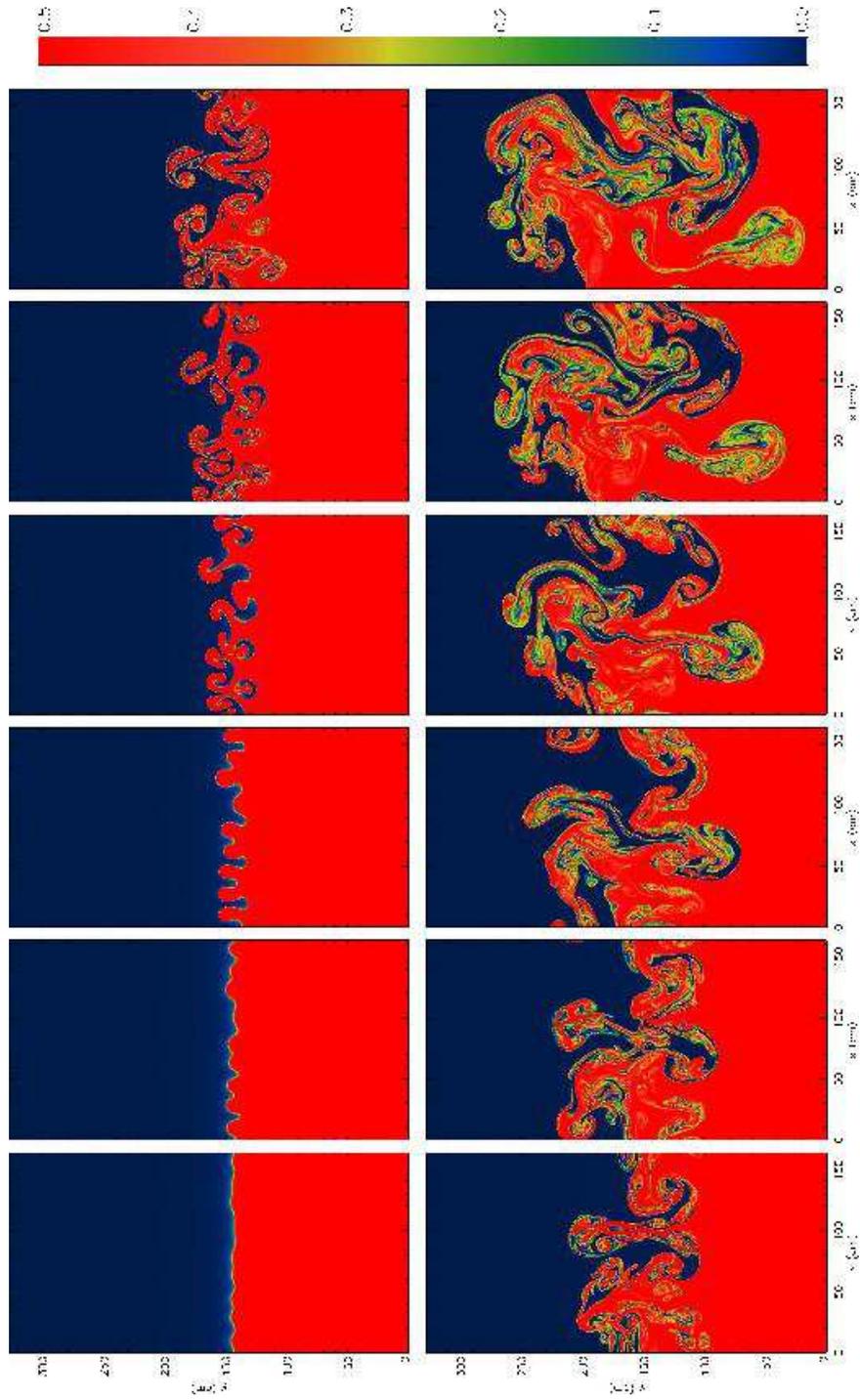}
\end{center}
\caption{\label{fig:rt_1.e7_noburn} Carbon mass fraction for
$10^7~\gcc$ C/O non-reactive RT simulation shown at the same times as
Figure~\ref{fig:rt_1.e7}.  Gravity points toward increasing~$y$.}
\end{figure*}

We can make fits of the growth of the mixed region to time by fitting to a 
power law, 
\begin{equation}
w(t) = c t^n \enskip .
\end{equation}
We do not attempt to fit to the Sharp-Wheeler scaling and find a value
of $\alpha$ because of the limited range of wavelengths we follow, and
the uncertainty as to whether that relation holds in the reactive
case.  We exclude the initial linear phase of the growth of the RT
instability in our fits.  Figures~\ref{fig:rt_6.67e6_width_fit} to
\ref{fig:rt_1.5e7_width_fit} shows the results.  In all cases, we find
$n$ less than 2.0, indicating that we are not in the Sharp-Wheeler
regime.  Furthermore, $n$ increases with increasing density.  We find
$n = 1.16$ for $6.67\times 10^6~\gcc$, $n = 1.2$ for $10^7~\gcc$, and
$n = 1.55$ for $1.5\times 10^7~\gcc$.  Larger scale studies, with a
greater range of wavelengths are needed to determine whether this is a
general result for reactive Rayleigh-Taylor, or if this is because of
the small size domains considered here.  One way we can
get some insight into this is to rerun one of these simulations
without burning.  (Note that in this case, we shift to a non-moving
frame of reference whereas the reacting cases are performed in a
reference frame moving at the laminar flame speed.)  The growth rate
for the $10^7~\gcc$ run with burning disabled is over-plotted in
Figure~\ref{fig:rt_1.e7_width}.  We see that it is much smoother than
the corresponding reactive case, since the spikes of fuel which push
into the hot ash never burn away.  The only wiggles in the mixing
region growth curves result from bubble mergers.  The fit to a power
law for this non-reactive simulation is presented in
Figure~\ref{fig:rt_1.e7_noburn_width_fit}.  The carbon mass fraction
for several points in time are shown in
Figure~\ref{fig:rt_1.e7_noburn}, which compares directly to
Figure~\ref{fig:rt_1.e7}.  Here, $n = 1.36$, higher than the reactive
case, but still not equal to 2.0.  The difference between this value
and the reactive case shows that the burning does have some influence.
In the non-reactive case we really do expect $n=2.0$, and therefore
the difference we see is very likely because we also are probably not
following enough modes to see the Sharp-Wheeler scaling.

\begin{figure*}
\begin{center}
\epsscale{0.7}
\plotone{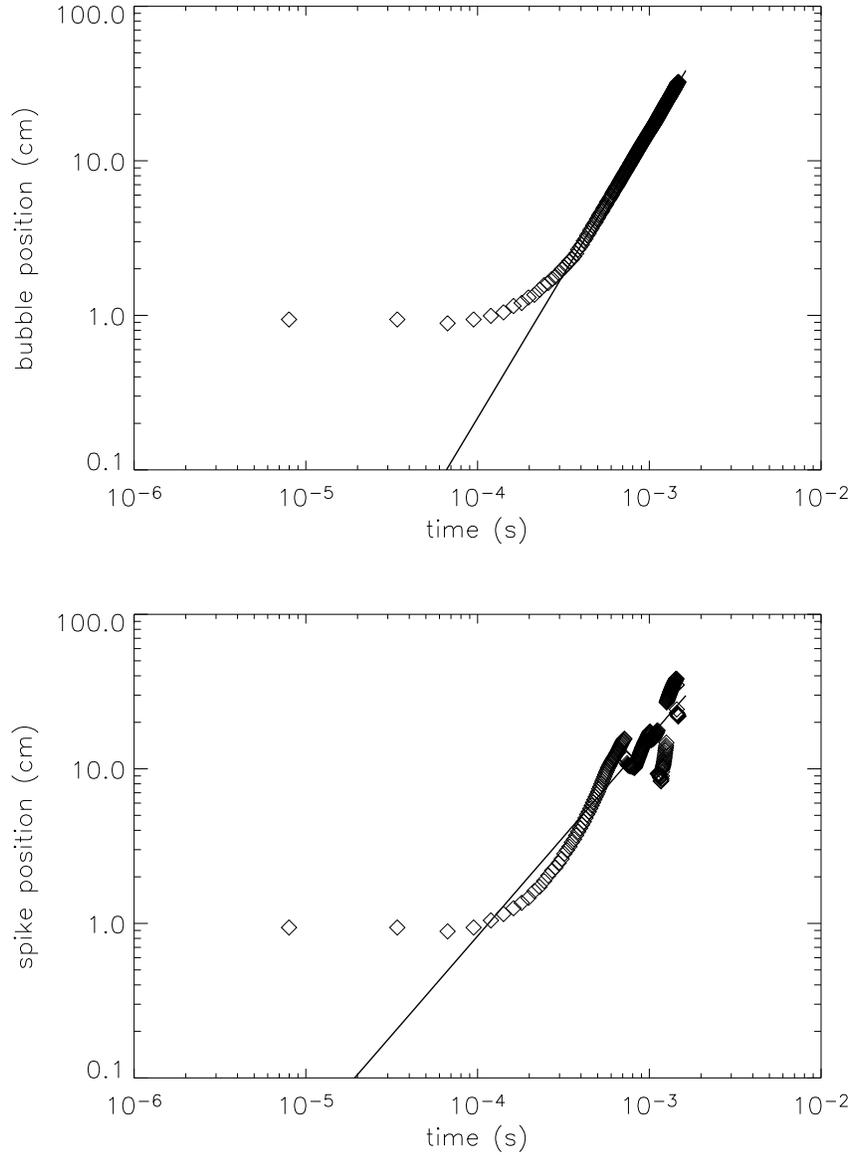}
\end{center}
\caption{\label{fig:rt_1.5e7_bubble_spike_fit} Fits of $c t^n$ to the
bubble and spike positions for the $1.5\times 10^7~\gcc$ RT run over
the time interval [$3\times 10^{-4}$~s, $1.45\times 10^{-3}$~s].  Here
we find the bubble position growing with $n=1.85$ and the spike region
growing much slower, with $n=1.28$.  This difference is to be expected
due to the strong influence of the reactions burning away the spikes
of fuel.}

\end{figure*}

For the $1.5\times 10^7~\gcc$ run, we were closest to growing as
$t^2$.  The burning here is quite vigorous, compared to the RT growth,
so as argued above, we would expect to see its influence on the extent
of the mixed region.  However, we would expect it to affect the
position of the spikes of fuel only, since the bubbles of ash do not
react.  Therefore, we can fit the growth of the bubbles and spikes
from the initial interface separately.  This is shown in
Figure~\ref{fig:rt_1.5e7_bubble_spike_fit}.  Again, we exclude the
initial linear growth phase of the instability.  We find the bubble
position growing as $n=1.85$ and the spike region growing as $n=1.28$.
Based on the trends we see here and above, we would expect that once
we have enough unstable modes in the box, the width of the mixed
region during the bubble merger phase will scale as slightly smaller
$t^2$, with the bubble position itself growing as $t^2$.  Therefore,
more studies, in larger domains with more unstable modes are needed.

\begin{figure*}
\begin{center}
\epsscale{0.5}
\plotone{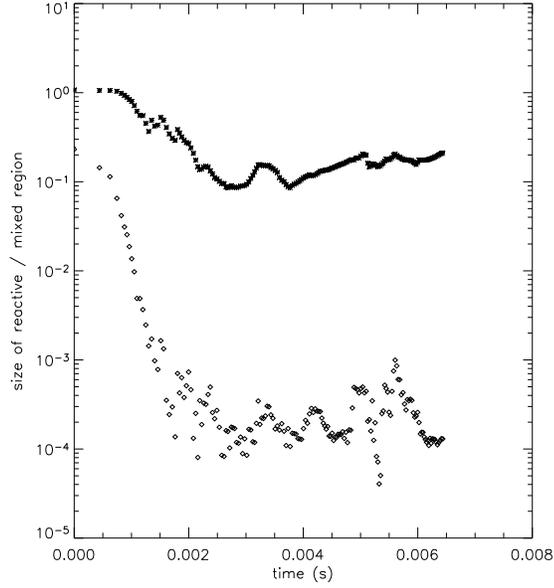}
\epsscale{1.0}
\end{center}
\caption{\label{fig:6.67e6_scale} Ratio of the size of the reactive
region to the mixed region, $\Gamma$ (see Equation~\ref{eq:gamma}), as a
function of time for the $6.67\times 10^6~\gcc$ RT unstable flame.
Two curves are shown, differing only in the choice of the energy
generation rate tolerance, $\gamma$ (see
Equation~\ref{eq:reactive_volume}).  The `$\times$' are $\gamma = 0.1$
and the `$\diamond$' are $\gamma = 0.8$.  After an initial transient,
the points appear to level off, indicating that the reactive region
grows in direct proportion to the size of the mixed region.  The large
separation between the two curves, resulting from the strongly peaked
nature of the nuclear energy generation rate, does not seem to affect
the trend.}
\end{figure*}

\begin{figure*}
\begin{center}
\epsscale{0.8}
\plotone{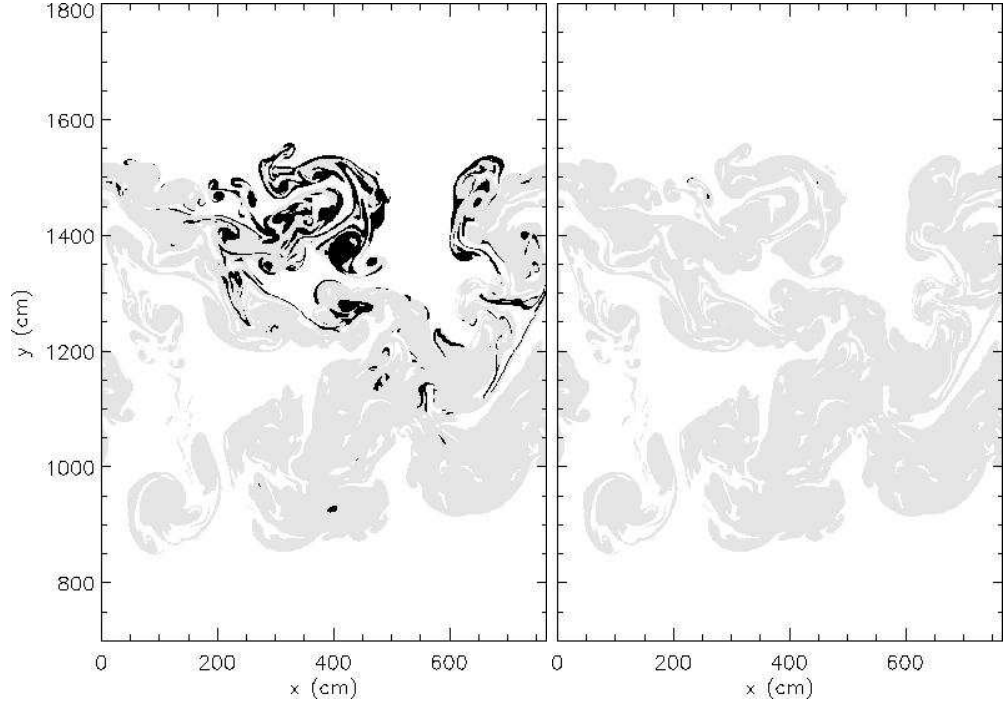}
\epsscale{1.0}
\end{center}
\caption{\label{fig:6.67e6_scale_thresh} Effect of the nuclear energy
generation rate threshold parameter, $\gamma$, on the relative sizes
of the reactive and mixed regions for the $6.67\times 10^6~\gcc$ RT
run at $4\times 10^{-3}$~s.  The mixed region, as defined by
Equation~\ref{eq:mixed_volume}, appears in gray and the reactive
region, Equation~\ref{eq:reactive_volume}, is in black, with $\gamma =
0.1$ (left) and $\gamma = 0.8$ (right).  Because the reaction rate is
strongly peaked, we see a large change in the size of the reactive
region between the two panes.  This is also reflected in
Figure~\ref{fig:carbon_destruction}.  Interestingly, regardless of
which value of $\gamma$ we use, the ratio of the volumes of the
reactive region to the mixed region appears to reach a steady state in
time (see Figure~\ref{fig:6.67e6_scale}).}
\end{figure*}

\begin{figure*}
\begin{center}
\epsscale{0.8}
\plotone{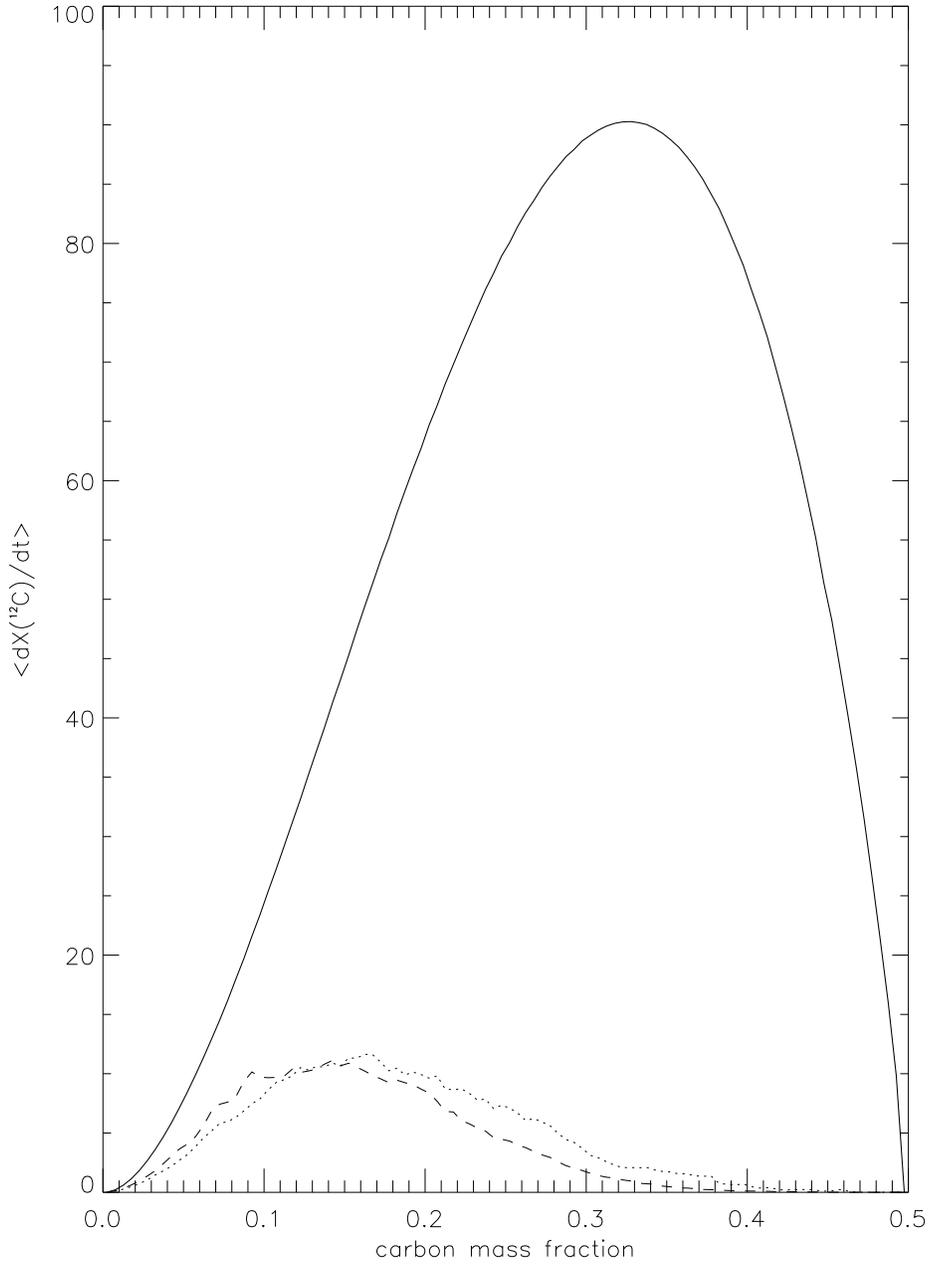}
\epsscale{1.0}
\end{center}
\caption{\label{fig:rt_6.67e6_yc12_dydt} Average carbon destruction
rate as a function of carbon mass fraction for the $6.67\times
10^6~\gcc$ simulation at $0$~s (solid), $3.2\times 10^{-3}$~s (dot),
and $6.4\times 10^{-3}$~s (dash).  The average was computed by binning
the mass fractions into 100 bins of width $5\times 10^{-3}$ and
computing the average value of $\mathrm{d}X_C/\mathrm{d}t$ for all zones
whose mass fraction falls in the bin.}
\end{figure*}

\begin{figure*}
\begin{center}
\epsscale{0.8}
\plotone{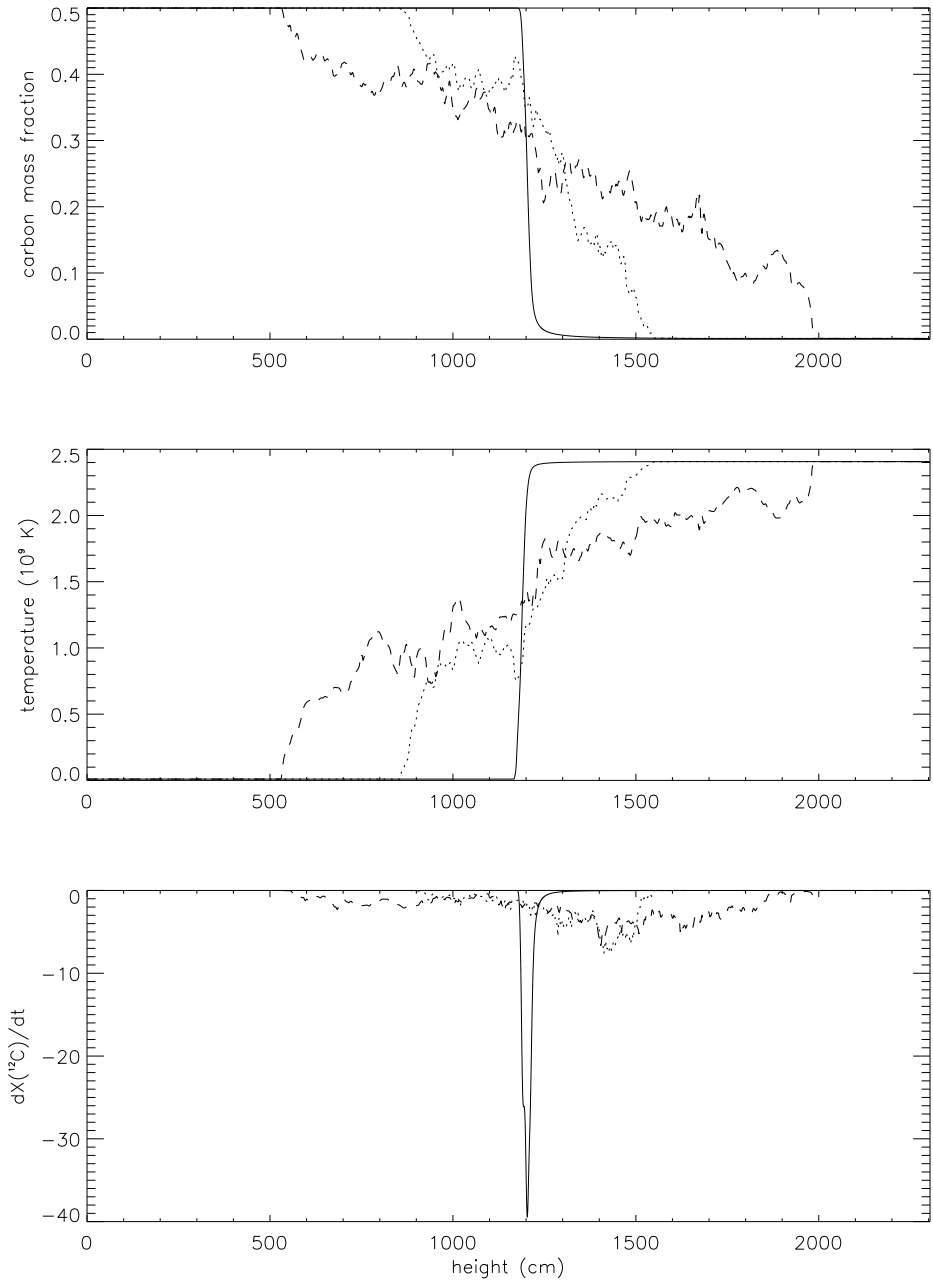}
\end{center}
\caption{\label{fig:rt_6.67e6_wide_averages} Laterally averaged
profiles for the $6.67\times 10^6~\gcc$, 768~cm wide, RT unstable
flame showing carbon mass fraction (top), temperature (middle), and
carbon destruction rate (bottom), for three times: $0$~s (solid),
$3.2\times 10^{-3}$~s (dot), and $6.4\times 10^{-3}$~s (dash).  We
notice that as time evolves, the `flame' becomes much broader,
and the burning rate sharply decreases.  In these
figures, the flame is moving to the left.}
\end{figure*}

It is interesting to look at how the size of the reactive region
scales with that of the mixed region, especially at the lowest
density.  As discussed in \S\ref{sec:rt}, if the reactive region grows
to the size of the detonation matchhead, then it may be possible for a
deflagration-detonation transition to occur.  Measuring the scaling of
the reactive region to the mixed region is not easy, because the flame
is so wrinkled.  We will use a volume weighted definition here.  The
collection of zones, $A$, where the carbon mass fraction,
$X_C$, is between $\eta_{\mathrm{min}}$ and
$\eta_{\mathrm{max}}$ is
\begin{equation}
\label{eq:mixed_volume}
A \equiv \left \{ X_{C\, (i,j)} \Bigm| 
           \eta_{\mathrm{min}} \le X_{C\, (i,j)} \le \eta_{\mathrm{max}} \right \} \enskip .
\end{equation}
Similarly, we can define the collection of zones, $B$, where the
density-weighted carbon destruction rate, is within $\gamma$ of the
peak,
\begin{equation}
\label{eq:reactive_volume}
B \equiv \left \{ \left ( \rho \dot{X}_C \right )_{(i,j)} \Bigm| 
           \left ( \rho \dot{X}_C \right )_{(i,j)} \le
	   \gamma  \max \left ( \left ( \rho \dot{X}_C \right )_{(i,j)} \right ) \right \} \enskip .
\end{equation}
Then, the ratio of the size of the reactive region to the mixed region
is
\begin{equation}
\label{eq:gamma}
\Gamma = \frac{V_B}{V_A} \enskip ,
\end{equation}
where $V_A$ is the volume of the zones in set $A$.  Once the mixed
region develops, we expect this quantity to be always less than unity.
Because of the freedom to choose the parameters $\eta_{\mathrm{min}}$,
$\eta_{\mathrm{max}}$, and $\gamma$, the numerical value of $\Gamma$
itself does not have much meaning, but the trend with time does.  We
pick $\eta_{\mathrm{min}} = 0.05$, $\eta_{\mathrm{max}} = 0.45$, and
two values of $\gamma$: 0.1 and 0.8.  We note that the trend is
relatively insensitive to the choice of these parameters, as
illustrated by Figure~\ref{fig:6.67e6_scale}, which shows $\Gamma$ as
a function of time for the $6.67\times 10^6~\gcc$ RT unstable flame
for the two choices of $\gamma$.  After an initial transient, it
appears that these curves level off, indicating that the reactive
region continues to growth in size proportional to the mixed region as
the flame evolves.  We note that this shows that the reactive region
is a small fraction of the mixed region, but this absolute scaling is
dependent on the choice of the three parameters.  This is illustrated
in Figure~\ref{fig:6.67e6_scale_thresh}, where the mixed and reactive
regions are shown, midway through the calculation.  At late times, the
peak nuclear energy generation comes from regions where the carbon
mass fraction is $\sim 0.15$, as shown in
Figure~\ref{fig:rt_6.67e6_yc12_dydt}.  This is also apparent by
looking at the laterally averaged flame profile and comparing to the
laminar state---see Figure~\ref{fig:rt_6.67e6_wide_averages}.  If this
reactive region can growth to a critical matchhead size, before the
entire star is consumed, it may be possible for a deflagration to
detonation transition to proceed.  However, since at late times the
peak burning is occurring at mass fractions of 0.15 rather than 0.5,
this will greatly increase the required matchhead size, further
complicating the possibility of a deflagration-detonation transition.

\subsection{\label{sec:res} Resolution and Domain Size Dependence on the Results}

\begin{figure*}
\begin{center}
\epsscale{0.8}
\plotone{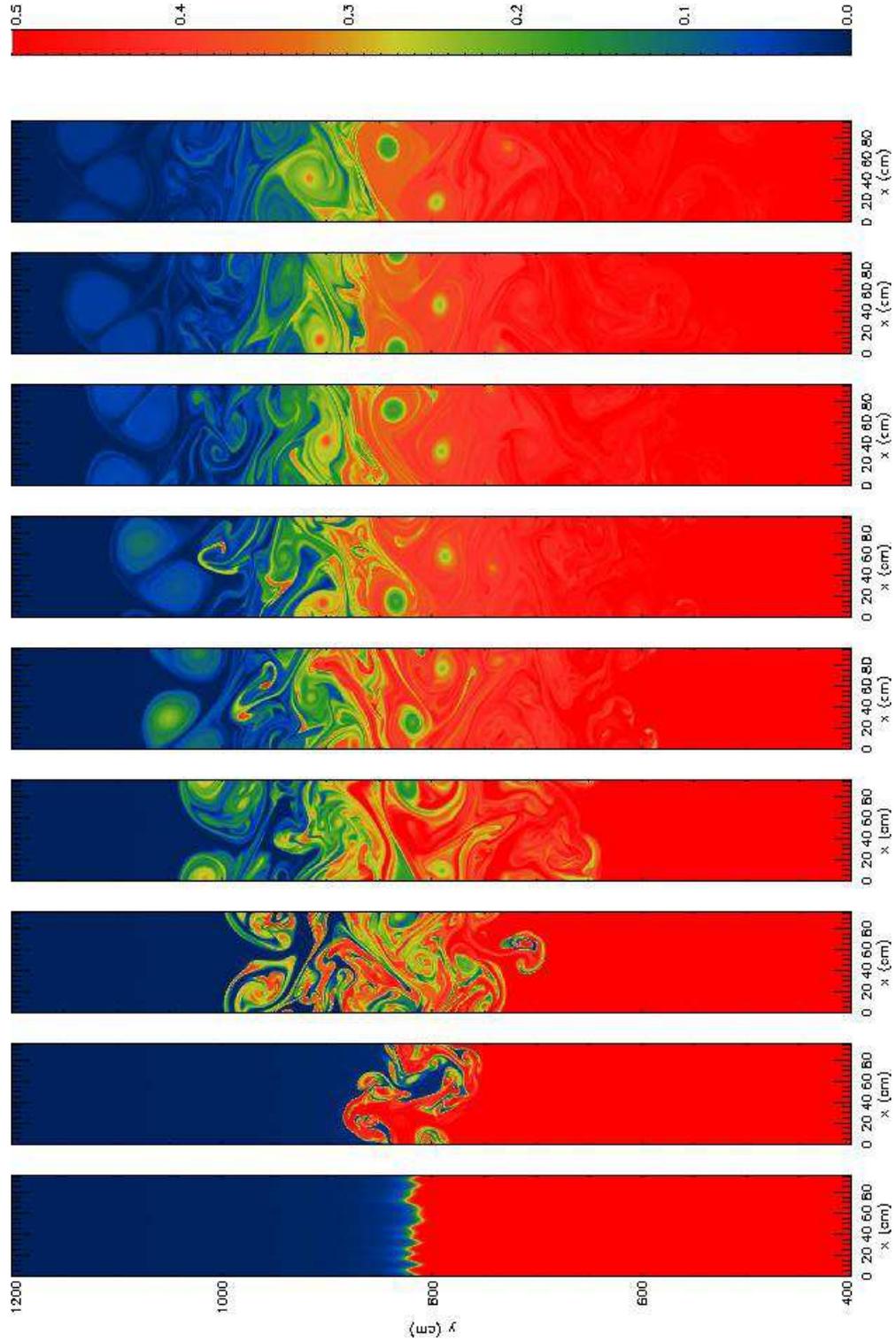}
\end{center}
\caption{\label{fig:rt_6.67e6} Carbon mass fraction for a 96~cm wide,
$6.67\times 10^6~\gcc$ C/O flame shown every $1.6\times 10^{-3}$~s,
starting at $0$~s, for the first $1.27\times 10^{-2}$~s of evolution.
The fuel (0.5~$^{12}$C by mass) appears red here and gravity points
up.  At late times, when the size of the mixed region becomes
comparable to the domain width, a steady state shear layer forms,
preventing the further growth of the instability.}
\end{figure*}

\begin{figure*}
\begin{center}
\epsscale{0.8}
\plotone{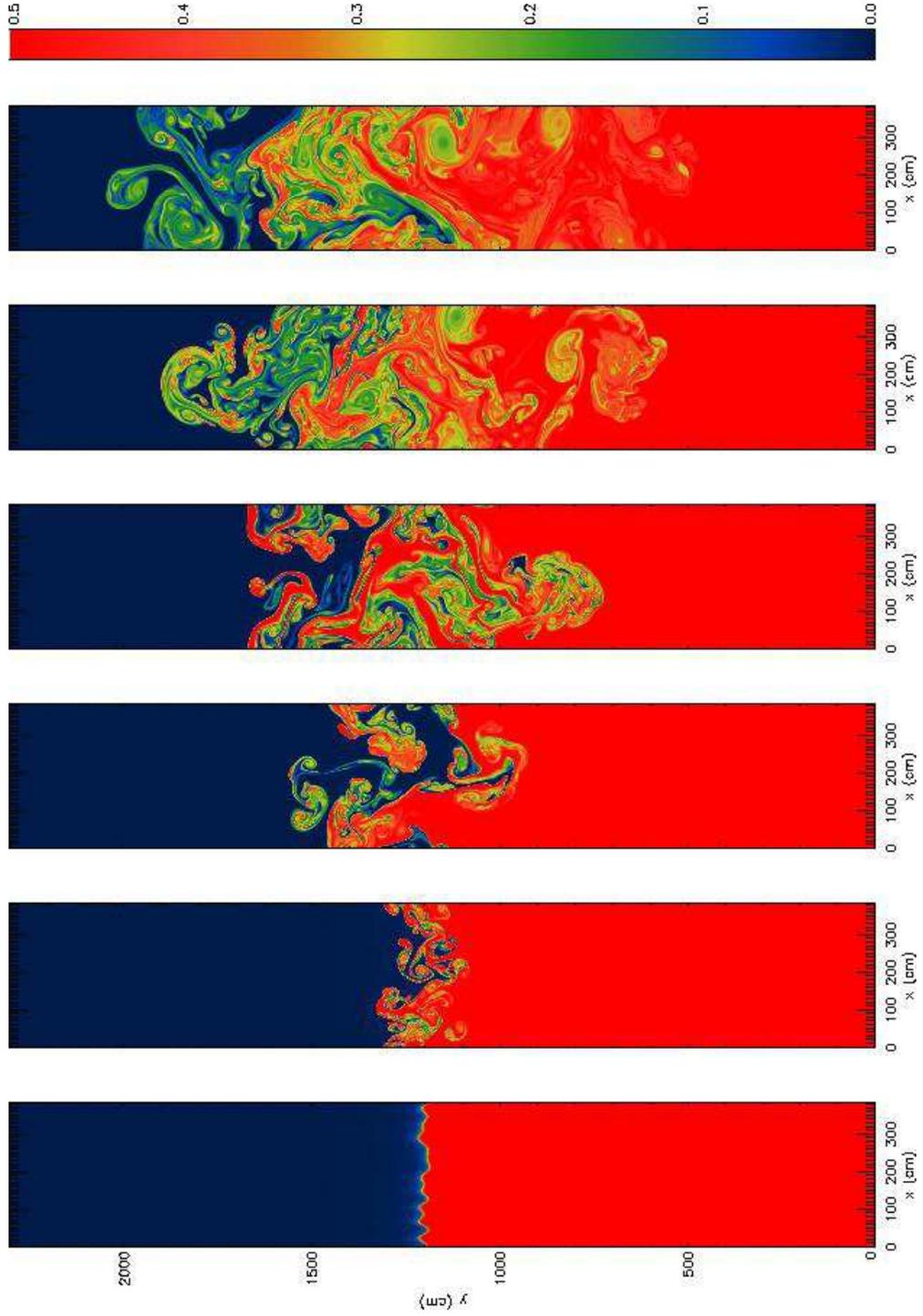}
\end{center}
\caption{\label{fig:rt_6.67e6_small2} Carbon mass fraction for a
384~cm wide, $6.67\times 10^6~\gcc$ C/O flame shown every $1.6\times
10^{-3}$~s until $8.12\times 10^{-2}$~s.  The fuel appears red (carbon
mass fraction = 0.5), and gravity points toward increasing~$y$.  At this low density, the
RT instability dominates over the burning, and a large mixed region
develops.}
\end{figure*}

In this subsection, we examine the role of domain size and resolution
on the computational results.  Figures~\ref{fig:rt_6.67e6}
and~\ref{fig:rt_6.67e6_small2} show the evolution of the $6.67\times
10^{6}~\gcc$ RT simulation in narrower domains (96~cm and 384~cm wide
respectively).  Other than the domain size, the parameters are
identical to those used in the results described shown in
Figure~\ref{fig:rt_6.67e6_wide}.  In the narrowest case, once the size
of the mixed region reaches a fixed size, comparable to the width of
the domain, the mixing and reacting processes saturate and the flow
enters a quasi-steady-state, with vortices that are remnants of the
early mixing driving transverse shear flow in the mixed region.  Once
this saturation is reached, the burning rate begins to slowly decay.
This behavior is not seen in the corresponding wider domain
calculation, although, we would expect a similar pattern if we use a
taller domain and ran out to longer times.


Figure~\ref{fig:rt_6.67e6_speeds} shows three additional curves
corresponding to these narrower domains, the narrowest at two
different resolutions (see Table~\ref{table:simparams}).  In the 96~cm
wide domain, the peak velocity is much smaller, reaching only about
3~times the laminar speed.  This is to be expected from
Equation~(\ref{eq:sw2})---the wider domain allows longer wavelength
modes to go unstable.  The decrease in velocity at late times in the
narrow domain begins when the size of the mixed layer becomes
comparable to the width of the domain, preventing any further modes
from growing.  This is reflected in the later panels of
Figure~\ref{fig:rt_6.67e6}.  In the 384~cm wide domain, the velocity
continues to grow, beyond 6~times the laminar speed---we would expect
our widest domain run to continue to accelerate as well, but it began
to interact with the top of our computational domain and was not run
out as long.

The velocity of the low resolution run tracks that of the high
resolution case very closely, suggesting that, for the large scale
diagnostics, we have converged.  This lowest resolution run has
approximately 5~points in the thermal width
(Equation~\ref{eq:thermalwidth}), a value found acceptable in the
convergence study performed in \citet{SNeCodePaper} for
Landau-Darrieus unstable astrophysical flames.  We note again that
this definition of thermal width is smaller than some other commonly
used definitions (see, for example \citealt{flame-curvature}).
Finally, in all calculations, the initial perturbations are well
resolved, with typically 50--100 zones per wavelength (with the only
exception being the 96~cm wide, $6.67\times 10^6~\gcc$ run, where
there are only 20 zones per wavelength).  This zoning is at or exceeds
the resolution determined to give acceptable convergence of
single-mode RT growth rates in the 3-d compressible study presented in
\citet{flash-validation}.

\begin{figure*}[H]
\begin{center}
\epsscale{0.5}
\plotone{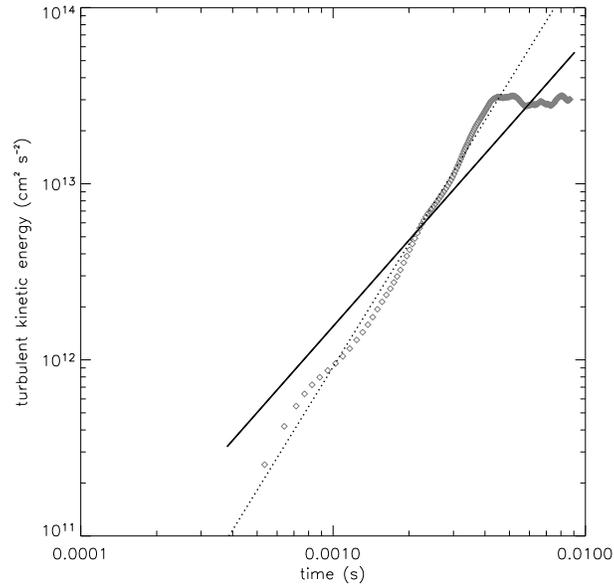}
\end{center}
\caption{\label{fig:rt_6.67e6_tke_small2} Turbulent kinetic energy as
a function of time for the 384~cm wide $6.67\times 10^6~\gcc$ RT
unstable flame.  The symbols are the data and two fits are provided.
The solid line is all the data, $k(t) = 1.169\times 10^{17}
t^{1.626}$.  The dotted line is just the initial $0.005$~s, the period
up to the point where the mixed region grows to rival the domain wide
in extent, and gives $k(t) = 8.602\times 10^{18} t^{2.325}$.
This latter fit agrees well with the wider domain results,
Figure~\ref{fig:rt_6.67e6_tke}.}
\end{figure*}

We also can look at how the scaling of the turbulent kinetic energy
depends on the domain size.  For the 384~cm wide, $6.67\times
10^6~\gcc$ run, this is shown in
Figure~\ref{fig:rt_6.67e6_tke_small2}.  Two fits are shown.  When all
of the data is included, the fit is quite poor, since it is biased by
by data at the end of the run when the mixing has saturated and the
turbulent kinetic energy stops increasing.  Excluding this data, the
fit to the first $0.005$~s is quite good, and with the kinetic energy
$\sim t^{2.325}$ quite close to the 2.267 power obtained from the
768~cm wide run.  Thus, it appears that the scaling of the turbulent
kinetic energy prior to saturation is insensitive to the domain size.

\section{DETONATION IN THE DISTRIBUTED REGIME?}
\label{sec:detonation}

Based upon the observed systematics of flame propagation in the
distributed regime, we can begin to speculate on the potential for a
transition to detonation.  As discussed by \citet{niemeyerwoosley1997}
and \citet{niemeyer1999}, there are two possible modes for a delayed
transition to detonation: a) a ``local'' transition to detonation
because a fluid element of critical mass burns with a supersonic phase
velocity and b) a ``macroscopic'' transition because some appreciable
fraction of the white dwarf volume develops such complex topology that
burning briefly consumes more fuel than could a spherical front
encompassing that region and moving at supersonic speed. The former,
also known as the Zeldovich mechanism, is the basis for the ``delayed
detonation model'' by \citet{khokhlov1991}; the latter is the basis
for a similar model by \citet{woosley1994}.  \citet{niemeyer1999} has
argued that the volume detonation model requires special
preconditioning, but because we study only the small scales on which
the flame is resolved, we cannot comment in a meaningful way on this
issue.  Also, because we have not included turbulence cascading down
from scales larger than our grid, we cannot conclusively argue about
the local transition to detonation either (a large amount of
turbulence could, for example, induce a transition to distributed
burning at a higher density). Still some scaling relations are
observed that, if they can be generalized by larger scale studies,
argue against a transition to detonation.

As is well recognized \citep[e.g.][]{niemeyer1999}, transition to
detonation can never occur so long as a well defined flame exists.
The width of a steady flame is always thinner than the critical mass
required for detonation.  However, a mixture of hot fuel and cold ash
is potentially explosive as calculations in the SN Ia context
\citep{lisewski2000} have suggested.  Figure
\ref{fig:carbon_destruction} shows the existence, in the distributed
regime, of localized hot spots of unsteady burning.  The high
temperature sensitivity of the $^{12}$C + $^{12}$C reaction itself
makes it difficult for one of these hot spots to become supercritical
in mass. For the conditions we consider, carbon fusion dominates the
energetics, and the energy production is
\begin{equation}
S_{\rm nuc} \ \propto \ X_C^2 \, T^n \enskip,
\end{equation}
where $X_C$ is the mass fraction of carbon, and $n$ given by the 
Coulomb barrier between two carbons nuclei,
\begin{equation}
n \ = \ \frac{\tau - 2}{3} \enskip ,
\end{equation}
\begin{eqnarray}
\tau  &=& 4.248 \, \left (\frac{Z_1^2 \, Z_2^2 \, \bar A}{T_9} \right)^{1/3}\\
&=&  84.13 \, T_9^{1/3},
\end{eqnarray}
with $T_9$ the temperature in 10$^9$ K. For our calculations with
$\rho$ = $6.67 \times 10^6~\gcc$, the temperature of the hot ash is
$T_9 = 2.4$. Anticipating that the temperature of a combustible
fuel-ash mixture will not be far from that, we find $n \approx 20$.
The carbon mass fraction in the fuel is 0.5 and in the ash it is zero,
and the temperature of the unburned fuel is negligible.

In any mixed mass, $M$, composed of fractions $f$ of ash and $1-f$ of
fuel, the temperature will be
\begin{equation}
T_{\rm mix} \ = \ f \, \frac{C_{\rm P}(T_{\rm ash})}{C_{\rm P}(T_{\rm mix})}
\ T_{\rm ash}
\end{equation}
and the carbon mass fraction
\begin{equation}
X_{C, {\rm mix}} \ \approx \ (1 - f) \, X_{C, {\rm fuel}} \enskip .
\end{equation}
The dependence of the heat capacity on temperature is important and
mitigates somewhat the extreme sensitivity of the reaction rate.  In
the vicinity of $T_9$ = 2.5, the heat capacity is predominantly due to
the electron gas (with $C_{\rm P,e} \propto T$), but with a
non-negligible contribution from radiation ($C_{\rm P} \propto T^3$)
\citep{woosley2003}. To good approximation, near $T_9$ = 2.5, $C_{\rm
P} \propto T^{7/4}$.
 
The maximum in energy generation will then occur for 
\begin{equation}
\frac{d \ }{d f} [(1 - f)^2 \ f^{80/11}] \ = \ 0 \enskip ,
\end{equation}
or for $f \approx 40/51 = 0.784$---that is, in the burning mixture,
$X_C = 0.11$ and $T_9 = 2.20$.  A factor of two variation in the
energy generation occurs for $0.60 < f< 0.91 $, or $X_C$ from $0.045$
to $0.20$.  This agrees well with what is observed in the numerical
simulation (Fig. \ref{fig:rt_6.67e6_yc12_dydt}).

Coupled with the fact that burning to magnesium or silicon releases
less energy than burning to the iron group, this low carbon mass
fraction implies that the overpressure from burning will be small in
the regions where ash and fuel are mixed. Burning from $T_9 = 2.2$
to~$2.4$, the typical ash temperature, will only raise the pressure by
8\%. Since this overpressure determines the temperature reached in a
detonation and the burning rate depends on this to a high power, the
critical mass for initiating a detonation will be large. The size will
certainly be greater than the distance a sound wave can travel during
the time it takes the mixture to burn. At $T_9$ = 2.2 and $X_C =
0.11$, the nuclear time scale is
\begin{equation}
\tau_{\rm T} \ \approx \ \frac{C_{\rm P}}{\dot S_{\rm nuc}} \enskip ,
\end{equation}
or about 0.07~s. The speed of sound is approximately 5000~km~s$^{-1}$.
Hence a critical mass for these conditions must be at least
300~km in size.

On the other hand, Figure \ref{fig:6.67e6_scale} shows that, in steady
state, the size of the reactive region is approximately a small
constant times the size of the mixed region. Even allowing a range of
burning time scales of a factor of 10 across the burning region, the
explosive mixture only has dimensions about 10\% that of the region
that has been mixed by the RT instability. Taking $0.05 g_{\rm eff}
t^2$ as an upper bound to that (see \S\ref{sec:growth}), and realizing
that the temperature will decline still further in a few tenths of a
second because of expansion, the largest mixed, potentially explosive
region is less than 5~km.  Thus, though a transition to detonation is
in principal possible, there simply may not be room, nor time enough
to develop one in an exploding white dwarf.

\section{CONCLUSIONS}
\label{sec:conclusions}

We presented direct numerical simulations of reactive RT instabilities
in conditions appropriate to the late stages of a Type~Ia supernova
explosion.  Fully resolving the flame frees us from the need to
specify any model parameters, such as the flame speed, Markstein
length, etc.  In the density range we consider, the flame transitions
from having a distinct, well defined interface at the high density end
to being a chaotic, well mixed burning region at the lower density
end, with a transition density of $\sim 10^7~\gcc$.  This transition
is expected based on the arguments presented in \S\ref{sec:rt}.
Furthermore, this would suggest that at the even higher densities
characteristic of the early stage of the explosion, the flame surface
continues to be well defined, consistent with the conclusions of
\citet{nbr1999} and the 2-d flame-model simulations at $10^8~\gcc$
presented in \citet{khokhlov1994}.  The effects of the burning were
further made concrete by presenting the $10^7~\gcc$ case both with and
without reactions.

At the lowest density, the mixed region becomes very large, although,
still smaller than the critical carbon detonation matchhead size
\citep{niemeyerwoosley1997}.  It appears that the reactive region
grows in direct proportion to the size of the mixed region, making it
possible in theory that such a transition can occur, however, at these
low densities, much of the star is already consumed, and, based on
scaling predictions from the present simulations, the time required
for the reactive region to grow to the matchhead size seems to be
longer than the total explosion time.  In effect, the flame runs out
of star before a critical mass can be built up.  Future simulations
will explore this in more depth.

In all cases presented here, the flames accelerated considerably,
reaching speeds between~2 and 6~times the laminar flame speed.  This
is significantly larger than accelerations seen in small-scale\slash
resolved Landau-Darrieus instability studies (paper~I).  The growth of
the flame surface appeared to be well described by a fractal model,
with a fractal dimension of $\sim 1.7$.  In the cases considered, the
effective flame speed was asymptotically proportional to the increase
in flame surface area but the constant of proportionality was about
0.1 for the low density cases and 0.3 for the higher density case.
Thus, the flame acceleration was considerably less than would be
predicted by the increase in flame surface area alone.  This
proportionality constant may approach unity as the density is
increased, because the smallest scale on which the RT instability can
bend the flame (the fire-polishing length) increases dramatically with
density, and therefore the magnitude of the curvature that the flame
experiences locally decreases.  We would expect that on larger domains
the flame would continue to accelerate, further supporting the already
widely held view that the RT instability provides most of the
acceleration to the flame in a pure deflagration Type~Ia SNe
explosion.

The growth of the mixed region for the reactive RT instability appears
to be slower than the non-reactive case.  The present results do not
seem to support the Sharp-Wheeler model, but larger scale studies are
needed, with more bubbles merging over a longer period of time to
better understand the growth of the mixed region.  Understanding the
growth of the mixed region is critical to providing an accurate
subgrid model.

Future studies will focus on the three-dimensional counterparts to the
flames we discussed here.  Turbulence behaves very differently in two
and three dimensions, so we expect to see some differences in the
evolution of the instability.  Numerical results indicate that the
growth rate of the pure RT instability is faster in 3-d than in 2-d
\citep{kane2000,flash-validation}.  Furthermore, in 3-d the surface to
volume ratio of the spikes of fuel is larger than in 2-d, exposing
more fuel to the hot ash, so we may expect them to burn away more
quickly.  We still expect, based on the arguments presented in
\S\ref{sec:rt} and the results of the simulations, that the flame will
undergoing the same transition in behavior as the density decreases.
Larger two-dimensional studies in the flamelet regime are also needed,
moving toward higher densities.  Capturing a greater range of scales
on the grid will allow for validation of the fractal scaling model and
a better understanding of the moderation of the Sharp-Wheeler
prediction for the growth of the mixed region.  Finally, these fully
resolved simulations can serve as the basis for testing various models
to represent the flame on subgrid scales, which become necessary to
address the larger scale Type~Ia physics.

\acknowledgements 

The authors thank F.~X. Timmes for making his equation of state and
conductivity routines available online.  Support for this work was
provided by the DOE grant No.\ DE-FC02-01ER41176 to the Supernova
Science Center/UCSC and the Applied Mathematics Program of the DOE
Office of Mathematics, Information, and Computational Sciences under
the U.S. Department of Energy under contract No.\ DE-AC03-76SF00098.
SEW acknowledges NASA Theory Award NAG5-12036.  Some calculations were
performed on the IBM SP (seaborg) at the National Energy Research
Scientific Computing Center, which is supported by the Office of
Science of the DOE under Contract No.\ DE-AC03-76SF00098, the IBM
Power4 (cheetah) at ORNL, sponsored by the Mathematical, Information,
and Computational Sciences Division; Office of Advanced Scientific
Computing Research; U.S. DOE, under Contract No.\ DE-AC05-00OR22725
with UT-Battelle, LLC, and the UCSC UpsAnd cluster supported by an NSF
MRI grant AST-0079757.


\clearpage

\begin{table*}
\begin{center}
\caption{\label{table:flameproperties} Properties of 0.5~$^{12}$C/0.5~$^{16}$O
flames. {color{red} is the $\rho$ in $\Delta \rho / \rho$ actually $\bar{\rho}$} }
\begin{tabular}{rdrddr}
\tableline
\tableline
\multicolumn{1}{c}{$\rho$} & \multicolumn{1}{c}{${\Delta \rho}/{\rho} $} & $v_{\mathrm{laminar}}$ & \multicolumn{1}{c}{$l_f$\tablenotemark{a}} & \multicolumn{1}{c}{$\lambda_{\mathrm{fp}}$\tablenotemark{b}} & \multicolumn{1}{c}{M} \\
\multicolumn{1}{c}{($\gcc$)} &  & ($\cms$) & \multicolumn{1}{c}{(cm)} & \multicolumn{1}{c}{(cm)} & \\
\tableline
$6.67\times 10^6$ & 0.529 & $1.04\times 10^3$ & 5.6 & 0.026 & $3.25\times10^{-6}$\\[2mm]
$10^7$ & 0.482 & $2.97\times 10^3$ & 1.9 & 0.23 & $8.49\times 10^{-6}$ \\[2mm]
$1.5\times 10^7$ & 0.436  &  $7.84\times 10^3$ & 0.54 & 1.8 & $2.06\times10^{-5}$\\
\tableline
\end{tabular}
\end{center}
\tablenotetext{a}{Equation~\ref{eq:thermalwidth}.}
\tablenotetext{b}{Equation~\ref{eq:firepolishing}, taking $g = 10^9~\mathrm{cm}~\mathrm{s}^{-2}$.}
\end{table*}

\begin{table*}
\begin{center}
\caption{\label{table:simparams} Parameters for the RT simulations.}
\begin{tabular}{rddddddl}
\tableline
\tableline
\multicolumn{1}{c}{$\rho$} & \multicolumn{1}{c}{width} & \multicolumn{1}{c}{height} & \multicolumn{1}{c}{$\Delta x_{\mathrm{fine}}$\tablenotemark{a}} & \multicolumn{1}{c}{width/$l_f$} & \multicolumn{1}{c}{width/$l_\mathrm{fp}$} & \multicolumn{1}{c}{$l_f/\Delta x_{\mathrm{fine}}$} & comments \\
\multicolumn{1}{c}{($\gcc$)} & \multicolumn{1}{c}{(cm)} & \multicolumn{1}{c}{(cm)} & \multicolumn{1}{c}{(cm)} \\
\tableline
$6.67\times 10^6$ & 96.0 & 3072.0 & 1.0 & 17.1 & 3290 & 5.6 & \nodata \\
                  & 96.0 & 1536.0 & 0.5 & 17.1 & 3290 & 11.2 & \nodata \\
                  & 384.0 & 2304.0 & 1.0 & 68.6 & 14800 & 5.6 & \nodata \\
                  & 768.0 & 2304.0 & 1.0 & 137 & 29500 & 5.6 & \nodata \\[2mm]
$10^7$            & 163.84 &  327.7 & 0.16 & 86.2 & 712 & 11.9 & \nodata\\
                  & 163.84 &  327.7 & 0.16 & 86.2 & 712 & 11.9 & no burning\\[2mm]
$1.5\times 10^7$ & 53.5 &  107.0 & 0.0522 & 99.1 & 29.7 & 10.3 & \nodata \\
\tableline
\end{tabular}
\end{center}
\tablenotetext{a}{$\Delta x_{\mathrm{fine}}$ is the zone width
on the finest mesh.  In all simulations, $\Delta x = \Delta y$.}
\end{table*}



\begin{table*}
\begin{center}
\caption{\label{table:tke} Turbulent kinetic energy fits to $k(t) = a t^b$.}
\begin{tabular}{rrrl}
\tableline
\tableline
\multicolumn{1}{c}{$\rho$} & \multicolumn{1}{c}{$a$} & \multicolumn{1}{c}{$b$} & comments \\
\multicolumn{1}{c}{($\gcc$)}\\
\tableline
{$6.67\times 10^6$} & $1.169\times 10^{17}$ & 1.626 & 384~cm wide; all data \\
                  & $8.602\times 10^{18}$ & 2.325 & 384~cm wide; initial 0.005~s \\
                  &  $6.514\times 10^{18}$ & 2.267 & 768~cm wide \\[2mm]

$10^7$            & $8.597\times 10^{18}$ & 2.591 & \nodata \\[2mm]

$1.5\times 10^7$  & $2.202\times 10^{19}$ & 2.867  & \nodata \\
\tableline
\end{tabular}
\end{center}

\end{table*}

\clearpage

\clearpage

\clearpage

\clearpage

\clearpage

\clearpage

\clearpage

\clearpage

\clearpage

\clearpage





\clearpage

\clearpage

\clearpage

%

\clearpage

\clearpage

\clearpage

\clearpage

\clearpage

%
%
%

\clearpage

\begin{thebibliography}{58}
\expandafter\ifx\csname natexlab\endcsname\relax\def\natexlab#1{#1}\fi

\bibitem[{Almgren {et~al.}(1998)Almgren, Bell, Colella, Howell, \&
  Welcome}]{AlmBelColHowWel98}
Almgren, A.~S., Bell, J.~B., Colella, P., Howell, L.~H., \& Welcome, M. 1998,
  J. Comput. Phys., 142, 1

\bibitem[{{Arnett} {et~al.}(1971){Arnett}, {Truran}, \& {Woosley}}]{arnett1971}
{Arnett}, W.~D., {Truran}, J.~W., \& {Woosley}, S.~E. 1971, \apj, 165, 87

\bibitem[{{Bell} {et~al.}(2003{\natexlab{a}}){Bell}, {Day}, {Rendleman},
  {Woosley}, \& {Zingale}}]{SNld}
{Bell}, J.~B., {Day}, M.~S., {Rendleman}, C.~A., {Woosley}, S.~E., \&
  {Zingale}, M.~A. 2003{\natexlab{a}}, Astrophysical Journal, submitted,
  paper~I

\bibitem[{{Bell} {et~al.}(2003{\natexlab{b}}){Bell}, {Day}, {Rendleman},
  {Woosley}, \& {Zingale}}]{SNeCodePaper}
---. 2003{\natexlab{b}}, Journal of Computational Physics, accepted

\bibitem[{{Calder} {et~al.}(2002){Calder}, {Fryxell}, {Plewa}, {Rosner},
  {Dursi}, {Weirs}, {Dupont}, {Robey}, {Kane}, {Remington}, {Drake}, {Dimonte},
  {Zingale}, {Timmes}, {Olson}, {Ricker}, {MacNeice}, \&
  {Tufo}}]{flash-validation}
{Calder}, A.~C., {Fryxell}, B., {Plewa}, T., {Rosner}, R., {Dursi}, L.~J.,
  {Weirs}, V.~G., {Dupont}, T., {Robey}, H.~F., {Kane}, J.~O., {Remington},
  B.~A., {Drake}, R.~P., {Dimonte}, G., {Zingale}, M., {Timmes}, F.~X.,
  {Olson}, K., {Ricker}, P., {MacNeice}, P., \& {Tufo}, H.~M. 2002, \apjs, 143,
  201

\bibitem[{Caughlan \& Fowler(1988)}]{caughlan-fowler:1988}
Caughlan, G.~R., \& Fowler, W.~A. 1988, Atomic Data and Nuclear Data Tables,
  40, 283, see also
  \url{http://www.phy.ornl.gov/astrophysics/data/cf88/index.html}

\bibitem[{{Chandrasekhar}(1981)}]{chandra}
{Chandrasekhar}, S. 1981, Hydrodynamic and Hydromagnetic Stability (New York:
  Dover)

\bibitem[{Colella \& Woodward(1984)}]{ppm}
Colella, P., \& Woodward, P.~R. 1984, J. Comp. Phys., 54, 174

\bibitem[{{Cook} \& {Dimotakis}(2001)}]{cook2001}
{Cook}, A.~W., \& {Dimotakis}, P.~E. 2001, J. Fluid Mech., 443, 69

\bibitem[{Darrieus(1938)}]{darrieus:1938}
Darrieus, G. 1938, in La Technique Moderne, France

\bibitem[{{Davies} \& {Taylor}(1950)}]{daviestaylor1950}
{Davies}, R.~M., \& {Taylor}, G. 1950, Proc. R. Soc. London A, 200, 375

\bibitem[{Day \& Bell(2000)}]{daybell00}
Day, M.~S., \& Bell, J.~B. 2000, Combust. Theory Modelling, 4, 535

\bibitem[{{Dimotakis} {et~al.}(1998){Dimotakis}, {Catrakis}, {Cook}, \&
  {Patton}}]{dimotakis1998}
{Dimotakis}, P.~E., {Catrakis}, H.~J., {Cook}, A.~W., \& {Patton}, J.~M. 1998,
  {Presented at the 2nd Monte Verita Colloquium on Fundamental Problematic
  Issues in Turbulence, 22-28 March 1998 (Ascona, Switzerland). GALCIT Report
  FM98-2}

\bibitem[{{Dursi} {et~al.}(2003){Dursi}, {Zingale}, {Calder}, {Fryxell},
  {Timmes}, {Vladimirova}, {Rosner}, {Caceres}, {Lamb}, {Olson}, {Ricker},
  {Riley}, {Siegel}, \& {Truran}}]{flame-curvature}
{Dursi}, L.~J., {Zingale}, M., {Calder}, A.~C., {Fryxell}, B., {Timmes}, F.~X.,
  {Vladimirova}, N., {Rosner}, R., {Caceres}, A., {Lamb}, D.~Q., {Olson}, K.,
  {Ricker}, P.~M., {Riley}, K., {Siegel}, A., \& {Truran}, J.~W. 2003, \apj,
  595, 955

\bibitem[{{Gamezo} {et~al.}(2003){Gamezo}, {Khokhlov}, {Oran}, {Chtchelkanova},
  \& {Rosenberg}}]{gamezo2003}
{Gamezo}, V.~N., {Khokhlov}, A.~M., {Oran}, E.~S., {Chtchelkanova}, A.~Y., \&
  {Rosenberg}, R.~O. 2003, Science, 299, 77

\bibitem[{{George} {et~al.}(2002){George}, {Glimm}, {Li}, \& {Xu}}]{george2002}
{George}, E., {Glimm}, J., {Li}, X.-L., \& {Xu}, Z.-L. 2002, PNAS, 99, 2587

\bibitem[{{Glimm} \& {Li}(1988)}]{glimmli1988}
{Glimm}, J., \& {Li}, X.~L. 1988, Phys. Fluids, 31, 2077

\bibitem[{{Hasegawa} {et~al.}(1996){Hasegawa}, {Nishihara}, \&
  {Sakagami}}]{hasegawa1996}
{Hasegawa}, S., {Nishihara}, K., \& {Sakagami}, H. 1996, Fractals, 4, 241

\bibitem[{Hillebrandt \& Niemeyer(2000)}]{hillebrandtniemeyer2000}
Hillebrandt, W., \& Niemeyer, J.~C. 2000, Annu. Rev. Astron. Astrophys, 38, 191

\bibitem[{{Hoeflich} \& {Khokhlov}(1996)}]{hoeflichkhokhlov1996}
{Hoeflich}, P., \& {Khokhlov}, A. 1996, \apj, 457, 500

\bibitem[{{Kane} {et~al.}(2000){Kane}, {Arnett}, {Remington}, {Glendinning},
  {Baz{\' a}n}, {M{\" u}ller}, {Fryxell}, \& {Teyssier}}]{kane2000}
{Kane}, J., {Arnett}, D., {Remington}, B.~A., {Glendinning}, S.~G., {Baz{\'
  a}n}, G., {M{\" u}ller}, E., {Fryxell}, B.~A., \& {Teyssier}, R. 2000, \apj,
  528, 989

\bibitem[{{Khokhlov}(1993)}]{khokhlov1993}
{Khokhlov}, A. 1993, \apjl, 419, L77+

\bibitem[{{Khokhlov}(1994)}]{khokhlov1994}
---. 1994, \apjl, 424, L115

\bibitem[{{Khokhlov}(1991)}]{khokhlov1991}
{Khokhlov}, A.~M. 1991, \aap, 245, 114

\bibitem[{{Khokhlov}(1995)}]{khokhlov1995}
---. 1995, \apj, 449, 695

\bibitem[{{Khokhlov} {et~al.}(1997){Khokhlov}, {Oran}, \&
  {Wheeler}}]{khokhlov1997}
{Khokhlov}, A.~M., {Oran}, E.~S., \& {Wheeler}, J.~C. 1997, \apj, 478, 678

\bibitem[{{Kolmogoroff} {et~al.}(1937){Kolmogoroff}, {Petrovsky}, \&
  {Piscounoff}}]{kpp}
{Kolmogoroff}, A., {Petrovsky}, I., \& {Piscounoff}, N. 1937, Bulletin de
  l'universit{\'e} d'{\'e}tat {\`a} Moscou (S{\'e}rie internationale), section
  A, 1, 1, reprinted and translated in P. Pelc{\'e}, Dynamics of Curved Fronts,
  Academic Press, Berkeley, CA, 1988

\bibitem[{Landau(1944)}]{landau:1944}
Landau, L.~D. 1944, Acta Physicochimica, USSR, 19, 77, reprinted in P.
  Pelc{\'e}, Dynamics of Curved Fronts, Academic Press, Berkeley, CA, 1988

\bibitem[{{Layzer}(1955)}]{layzer1955}
{Layzer}, D. 1955, \apj, 122, 1

\bibitem[{{Lisewski} {et~al.}(2000){Lisewski}, {Hillebrandt}, {Woosley},
  {Niemeyer}, \& {Kerstein}}]{lisewski2000}
{Lisewski}, A.~M., {Hillebrandt}, W., {Woosley}, S.~E., {Niemeyer}, J.~C., \&
  {Kerstein}, A.~R. 2000, \apj, 537, 405

\bibitem[{{Livne}(1993)}]{Livne1993}
{Livne}, E. 1993, \apjl, 406, L17

\bibitem[{Majda \& Sethian(1985)}]{MajSet85}
Majda, A., \& Sethian, J.~A. 1985, Combust. Sci. Technol., 42, 185

\bibitem[{{M\"uller} \& {Arnett}(1982)}]{mullerarnett1982}
{M\"uller}, E., \& {Arnett}, W.~D. 1982, \apjl, 261, L109

\bibitem[{{M\"uller} \& {Arnett}(1986)}]{mullerarnett1986}
---. 1986, \apj, 307, 619

\bibitem[{{Niemeyer}(1999)}]{niemeyer1999}
{Niemeyer}, J.~C. 1999, \apjl, 523, L57

\bibitem[{Niemeyer {et~al.}(1999)Niemeyer, Busche, \& Ruetsch}]{nbr1999}
Niemeyer, J.~C., Busche, W.~K., \& Ruetsch, G.~R. 1999, \apj, 524, 290

\bibitem[{{Niemeyer} \&
  {Hillebrandt}(1995{\natexlab{a}})}]{niemeyerhillebrandt1995a}
{Niemeyer}, J.~C., \& {Hillebrandt}, W. 1995{\natexlab{a}}, \apj, 452, 779

\bibitem[{{Niemeyer} \&
  {Hillebrandt}(1995{\natexlab{b}})}]{niemeyerhillebrandt1995b}
---. 1995{\natexlab{b}}, \apj, 452, 769

\bibitem[{{Niemeyer} \& {Kerstein}(1997)}]{niemeyerkerstein1997}
{Niemeyer}, J.~C., \& {Kerstein}, A.~R. 1997, New Astronomy, 2, 239

\bibitem[{{Niemeyer} \& {Woosley}(1997)}]{niemeyerwoosley1997}
{Niemeyer}, J.~C., \& {Woosley}, S.~E. 1997, \apj, 475, 740

\bibitem[{{Nomoto} {et~al.}(1984){Nomoto}, {Thielemann}, \& {Yokoi}}]{nomoto84}
{Nomoto}, K., {Thielemann}, F.-K., \& {Yokoi}, K. 1984, \apj, 286, 644

\bibitem[{{O'Rourke} \& {Bracco}(1979)}]{orourke1979}
{O'Rourke}, P.~J., \& {Bracco}, F.~V. 1979, J. Comput. Phys., 33, 185

\bibitem[{Peters(2000)}]{peters:2000}
Peters, N. 2000, Turbulent combustion (Cambridge University Press)

\bibitem[{{Reinecke} {et~al.}(1999{\natexlab{a}}){Reinecke}, {Hillebrandt}, \&
  {Niemeyer}}]{reinecke1999b}
{Reinecke}, M., {Hillebrandt}, W., \& {Niemeyer}, J.~C. 1999{\natexlab{a}},
  \aap, 347, 739

\bibitem[{{Reinecke} {et~al.}(2002){Reinecke}, {Hillebrandt}, \&
  {Niemeyer}}]{reinecke2002b}
---. 2002, \aap, 391, 1167

\bibitem[{{Reinecke} {et~al.}(1999{\natexlab{b}}){Reinecke}, {Hillebrandt},
  {Niemeyer}, {Klein}, \& {Gr{\" o}bl}}]{reinecke1999a}
{Reinecke}, M., {Hillebrandt}, W., {Niemeyer}, J.~C., {Klein}, R., \& {Gr{\"
  o}bl}, A. 1999{\natexlab{b}}, \aap, 347, 724

\bibitem[{{Sharp}(1984)}]{sharp1984}
{Sharp}, D.~H. 1984, Physica, 12D, 3

\bibitem[{{Taylor}(1950)}]{taylor1950}
{Taylor}, G. 1950, Proc. R. Soc. London A, 201, 192

\bibitem[{Timmes(2000)}]{timmes_he_flames:2000}
Timmes, F.~X. 2000, \apj, 528, 913

\bibitem[{Timmes \& Swesty(2000)}]{timmes_swesty:2000}
Timmes, F.~X., \& Swesty, F.~D. 2000, \apjs, 126, 501

\bibitem[{Timmes \& Woosley(1992)}]{timmeswoosley1992}
Timmes, F.~X., \& Woosley, S.~E. 1992, \apj, 396, 649

\bibitem[{{Vladimirova} \& {Rosner}(2003)}]{vladimirova2003}
{Vladimirova}, N., \& {Rosner}, R. 2003, Phys. Rev. E, 67, 066305

\bibitem[{{Woosley}(1990)}]{woosley1990}
{Woosley}, S.~E. 1990, in Supernova, ed. A.~G. Petschek (New York:
  Springer-Verlag), 182

\bibitem[{{Woosley} {et~al.}(1984){Woosley}, {Axelrod}, \&
  {Weaver}}]{woosley84}
{Woosley}, S.~E., {Axelrod}, T.~S., \& {Weaver}, T.~A. 1984, in ASSL Vol. 109:
  Stellar Nucleosynthesis, 263

\bibitem[{{Woosley} \& {Weaver}(1994)}]{woosley1994}
{Woosley}, S.~E., \& {Weaver}, T.~A. 1994, in Supernovae, Les Houches, Session
  LIV, ed. S.~A. {Blundman}, R.~{Mochkovitch}, \& J.~{Zinn-Justin} (Amsterdam:
  North-Holland), 63--154

\bibitem[{{Woosley} {et~al.}(2003){Woosley}, {Wunsch}, \&
  {Kuhlen}}]{woosley2003}
{Woosley}, S.~E., {Wunsch}, S., \& {Kuhlen}, M. 2003, \apj, submitted,
  astro-ph/0307565

\bibitem[{{Young} {et~al.}(2001){Young}, {Tufo}, {Dubey}, \&
  {Rosner}}]{young2001}
{Young}, Y.-N., {Tufo}, H.~M., {Dubey}, A., \& {Rosner}, R. 2001, J. Fluid
  Mech., 447, 377

\bibitem[{{Zingale} {et~al.}(2001){Zingale}, {Niemeyer}, {Timmes}, {Duris},
  {Calder}, {Fryxell}, {Lamb}, {MacNeice}, {Olson}, {Ricker}, {Rosner},
  {Truran}, \& {Tufo}}]{flamevortex}
{Zingale}, M., {Niemeyer}, J.~C., {Timmes}, F.~X., {Duris}, L.~J., {Calder},
  A.~C., {Fryxell}, B., {Lamb}, D.~Q., {MacNeice}, P., {Olson}, K., {Ricker},
  P., {Rosner}, R., {Truran}, J.~W., \& {Tufo}, H.~M. 2001, in Relativistic
  Astrophysics: 20th Texas Symposium, 490

\end{thebibliography}
\end{document}